\documentclass[12pt]{article}

\usepackage{axodraw}
\usepackage{epsf}
\usepackage{rotating}

\newcommand{\beq}{\begin{equation}}
\newcommand{\eeq}{\end{equation}}
\newcommand{\bea}{\begin{eqnarray}}
\newcommand{\eea}{\end{eqnarray}}
\newcommand{\bed}{\begin{displaymath}}
\newcommand{\eed}{\end{displaymath}}

\newcommand{\non}{\nonumber}

\newcommand{\tgb}{{\rm tg}\beta}

\newcommand{\lsim}{\raisebox{-0.13cm}{~\shortstack{$<$ \\[-0.07cm] $\sim$}}~}

\catcode`\@=11

\def\@citex[#1]#2{\if@filesw\immediate\write\@auxout{\string\citation{#2}}\fi
  \def\@citea{}\@cite{\@for\@citeb:=#2\do
    {\@citea\def\@citea{,\penalty\@m}\@ifundefined
       {b@\@citeb}{{\bf ?}\@warning
       {Citation `\@citeb' on page \thepage \space undefined}}%
\hbox{\csname b@\@citeb\endcsname}}}{#1}}

\def\citer{\@ifnextchar [{\@tempswatrue\@citexr}{\@tempswafalse\@citexr[]}}
\def\@citexr[#1]#2{\if@filesw\immediate\write\@auxout{\string\citation{#2}}\fi
  \def\@citea{}\@cite{\@for\@citeb:=#2\do
    {\@citea\def\@citea{--\penalty\@m}\@ifundefined
       {b@\@citeb}{{\bf ?}\@warning
       {Citation `\@citeb' on page \thepage \space undefined}}%
\hbox{\csname b@\@citeb\endcsname}}}{#1}}
\catcode`\@=12

\begin{document}

\renewcommand{\thefootnote}{\fnsymbol{footnote}}
\setcounter{page}{0}

\begin{titlepage}

\vskip-1.0cm

\begin{flushright}
PSI--PR--06--15 \\
CERN--PH--TH/2006--265 \\
LAPTH--1168/06 \\
hep-ph/0612254
\end{flushright}

\begin{center}
{\large \sc Higgs Boson Production via Gluon Fusion:} \\[0.5cm]
{\large \sc Squark Loops at NLO QCD}\footnote{Supported in part by the
Swiss Bundesamt f\"ur Bildung und Wissenschaft.}\\
\end{center}

\vskip 1.cm
\begin{center}
{\sc Margarete M\"uhlleitner$^{1,2}$ and Michael Spira$^3$}

\vskip 0.8cm

\begin{small} 
{\it \small
$^1$ CERN, Theory Division, CH--1211 Gen\`eve 23, Switzerland \\
$^2$ Laboratoire d'Annecy-Le-Vieux de Physique Th\'eorique, LAPTH,
Annecy-Le-Vieux, France \\
$^3$ Paul Scherrer Institut, CH--5232 Villigen PSI, Switzerland}
\end{small}
\end{center}

\vskip 2cm

\begin{abstract}
\noindent
The loop-induced processes $gg\to h,H,A$ provide the dominant Higgs
boson production mechanisms at the Tevatron and LHC in a large range of
the minimal supersymmetric extension of the Standard Model. For squark
masses below $\sim 400$ GeV squark loop contributions become important
in addition to the top and bottom quark loops.  The next-to-leading
order QCD corrections to the squark contributions of these processes are
determined including the full squark and Higgs mass dependences. They
turn out to be large and thus important for the Tevatron
and LHC experiments. Squark mass effects of the $K$ factors can be of
${\cal O}(20\%)$. In addition we derive the QCD corrections to the
squark contributions of the rare photonic Higgs decays $h,H\to
\gamma\gamma$, which play a role for the Higgs searches at the LHC.
\end{abstract}

\end{titlepage}

\renewcommand{\thefootnote}{\arabic{footnote}}

\setcounter{footnote}{0}

\section{Introduction}
Higgs boson \cite{hi64} searches belong to the major endeavors at
present and future colliders within the Standard Model (SM) and its
minimal supersymmetric extension (MSSM). In the MSSM two isospin Higgs
doublets are introduced in order to generate masses of up- and down-type
fermions \cite{twoiso}. After electroweak symmetry breaking three of the
eight degrees of freedom are absorbed by the $Z$ and $W$ gauge bosons,
leaving five states as elementary Higgs particles. These consist of two
CP-even neutral (scalar) particles $h,H$, one CP-odd neutral
(pseudoscalar) particle $A$ and two charged bosons $H^\pm$.  At leading
order the MSSM Higgs sector is fixed by two independent input parameters
which are usually chosen as the pseudoscalar Higgs mass $M_A$ and
$\tgb=v_2/v_1$, the ratio of the two vacuum expectation values.
Including the one-loop and dominant two-loop corrections the upper bound
of the light scalar Higgs mass is $M_h\lsim 140$ GeV \cite{mssmrad}. The
couplings of the various neutral Higgs bosons to fermions and gauge
bosons, normalized to the SM Higgs couplings, are listed in
Table~\ref{tb:hcoup}, where the angle $\alpha$ denotes the mixing angle
of the scalar Higgs bosons $h,H$. An important property of the bottom
Yukawa couplings is their enhancement for large values of $\tgb$, while
the top Yukawa couplings are suppressed for large $\tgb$
\cite{schladming}.
\begin{table}[hbt]
\renewcommand{\arraystretch}{1.5}
\begin{center}
\begin{tabular}{|lc||ccc|} \hline
\multicolumn{2}{|c||}{$\phi$} & $g^\phi_u$ & $g^\phi_d$ &  $g^\phi_V$ \\
\hline \hline
SM~ & $H$ & 1 & 1 & 1 \\ \hline
MSSM~ & $h$ & $\cos\alpha/\sin\beta$ & $-\sin\alpha/\cos\beta$ &
$\sin(\beta-\alpha)$ \\ & $H$ & $\sin\alpha/\sin\beta$ &
$\cos\alpha/\cos\beta$ & $\cos(\beta-\alpha)$ \\
& $A$ & $ 1/\tgb$ & $\tgb$ & 0 \\ \hline
\end{tabular}
\renewcommand{\arraystretch}{1.2}
\caption[]{\label{tb:hcoup} \it Higgs couplings in the MSSM to fermions
and gauge bosons [$V=W,Z$] relative to the SM couplings.}
\end{center}
\end{table}

For this work we need the Higgs couplings to stop and sbottom squarks in
addition. The scalar superpartners $\tilde f_{L,R}$ of the left- and
right-handed fermion components mix with each other. The mass
eigenstates $\tilde f_{1,2}$ of the sfermions are related to the current
eigenstates $\tilde f_{L,R}$ by mixing angles $\theta_f$,
\begin{eqnarray}
\tilde f_1 & = & \tilde f_L \cos\theta_f + \tilde f_R \sin \theta_f
\nonumber \\
\tilde f_2 & = & -\tilde f_L\sin\theta_f + \tilde f_R \cos \theta_f \, .
\label{eq:sfmix}
\end{eqnarray}
The mass matrix of the sfermions in the left-right-basis is given by
\cite{mssmbase}\footnote{For simplicity, the $D$-terms have been
absorbed in the sfermion-mass parameters $M_{\tilde f_{L/R}}^2$.}
\begin{equation}
{\cal M}_{\tilde f} = \left[ \begin{array}{cc}
M_{\tilde f_L}^2 + m_f^2 & m_f (A_f-\mu r_f) \\
m_f (A_f-\mu r_f) & M_{\tilde f_R}^2 + m_f^2
\end{array} \right] \, ,
\label{eq:sqmassmat}
\end{equation}
with the parameters $r_d = 1/r_u = \tgb$ for down- and up-type fermions.
The parameters $A_f$ denote the trilinear scalar couplings of the soft
supersymmetry breaking part of the Lagrangian, $\mu$ the Higgsino mass
parameter and $m_f$ the fermion mass. The mixing angles acquire the form
\begin{equation}
\sin 2\theta_f = \frac{2m_f (A_f-\mu r_f)}{M_{\tilde f_1}^2 - M_{\tilde
f_2}^2}
~~~,~~~
\cos 2\theta_f = \frac{M_{\tilde f_L}^2 - M_{\tilde f_R}^2}{M_{\tilde
f_1}^2
- M_{\tilde f_2}^2} \, ,
\end{equation}
and the masses of the squark mass eigenstates are given by
\begin{equation}
M_{\tilde f_{1,2}}^2 = m_f^2 + \frac{1}{2}\left[ M_{\tilde f_L}^2 +
M_{\tilde f_R}^2 \mp \sqrt{(M_{\tilde f_L}^2 - M_{\tilde f_R}^2)^2 +
4m_f^2 (A_f - \mu r_f)^2} \right] \, .
\end{equation}
Since the mixing angles are proportional to the masses of the ordinary
fermions, mixing effects are only important for the third-generation
sfermions. The neutral Higgs couplings to sfermions read \cite{DSUSY}
\begin{eqnarray}
g_{\tilde f_L \tilde f_L}^\phi & = & m_f^2 g_1^\phi + M_Z^2 (I_{3f}
- e_f\sin^2\theta_W) g_2^\phi \nonumber \\
g_{\tilde f_R \tilde f_R}^\phi & = & m_f^2 g_1^\phi + M_Z^2
e_f\sin^2\theta_W
g_2^\phi \nonumber \\
g_{\tilde f_L \tilde f_R}^\phi & = & -\frac{m_f}{2} (\mu g_3^\phi
- A_f g_4^\phi) \, ,
\label{eq:hsfcouprl}
\end{eqnarray}
with the couplings $g_i^\phi$ listed in Table \ref{tb:hsfcoup}. $I_{3f}$
denotes the third component of the electroweak isospin, $e_f$ the
electric charge of the fermion $f$, $\theta_W$ the Weinberg angle and
$M_Z$ the $Z$-boson mass.  All these couplings have to be rotated to the
mass eigenstates by the mixing angle $\theta_f$.
\begin{table}[hbt]
\renewcommand{\arraystretch}{1.5}
\begin{center}
\begin{tabular}{|l|c||c|c|c|c|} \hline
$\tilde f$ & $\phi$ & $g^\phi_1$ & $g^\phi_2$ & $g^\phi_3$ & $g^\phi_4$
\\
\hline \hline
& $h$ & $\cos\alpha/\sin\beta$ & $-\sin(\alpha+\beta)$ &
$-\sin\alpha/\sin\beta$ & $\cos\alpha/\sin\beta$ \\
$\tilde u$ & $H$ & $\sin\alpha/\sin\beta$ & $\cos(\alpha+\beta)$ &
$\cos\alpha/\sin\beta$ & $\sin\alpha/\sin\beta$ \\
& $A$ & 0 & 0 & $-1$ & $1/\tgb$ \\ \hline
& $h$ & $-\sin\alpha/\cos\beta$ & $-\sin(\alpha+\beta)$ &
$\cos\alpha/\cos\beta$ & $-\sin\alpha/\cos\beta$ \\
$\tilde d$ & $H$ & $\cos\alpha/\cos\beta$ & $\cos(\alpha+\beta)$ &
$\sin\alpha/\cos\beta$ & $\cos\alpha/\cos\beta$ \\
& $A$ & 0 & 0 & $-1$ & $\tgb$ \\ \hline
\end{tabular}
\renewcommand{\arraystretch}{1.2}
\caption[]{\label{tb:hsfcoup}
\it Coefficients of the neutral MSSM Higgs couplings to up- and
down-type sfermion pairs.}
\end{center}
\end{table}

\begin{figure}[htb]
\begin{center}
\begin{picture}(130,90)(20,0)
\Gluon(10,20)(50,20){-3}{4}
\Gluon(10,80)(50,80){3}{4}
\ArrowLine(50,20)(50,80)
\ArrowLine(50,80)(90,50)
\ArrowLine(90,50)(50,20)
\DashLine(90,50)(130,50){5}
\put(0,18){$g$}
\put(0,78){$g$}
\put(30,46){$t,b$}
\put(100,36){$h,H,A$}
\end{picture}
\begin{picture}(130,90)(0,0)
\Gluon(10,20)(50,20){-3}{4}
\Gluon(10,80)(50,80){3}{4}
\DashLine(50,20)(50,80){5}
\DashLine(50,80)(90,50){5}
\DashLine(90,50)(50,20){5}
\DashLine(90,50)(130,50){5}
\put(0,18){$g$}
\put(0,78){$g$}
\put(30,46){$\tilde t,\tilde b$}
\put(100,36){$h,H$}
\end{picture}
\begin{picture}(130,90)(-20,0)
\Gluon(10,20)(50,50){-3}{5}
\Gluon(10,80)(50,50){3}{5}
\DashCArc(70,50)(20,180,360){5}
\DashCArc(70,50)(20,0,180){5}
\DashLine(90,50)(130,50){5}
\put(0,18){$g$}
\put(0,78){$g$}
\put(60,76){$\tilde t,\tilde b$}
\put(100,36){$h,H$}
\end{picture}  \\
\caption{\label{fg:lodiahgg} \it MSSM Higgs boson production via gluon
fusion mediated by top- and bottom quark as well as stop and sbottom
loops at leading order.}
\end{center}
\end{figure}
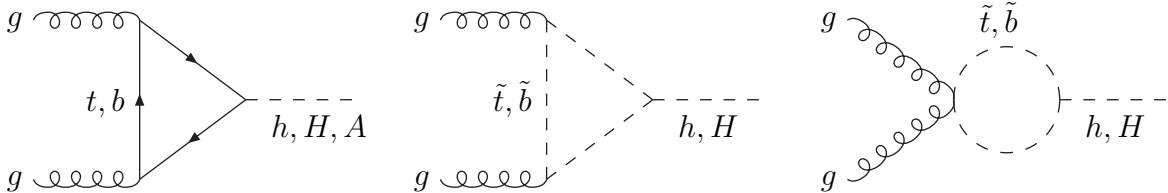

At hadron colliders as the Tevatron and LHC neutral Higgs bosons are copiously
produced by the gluon fusion processes $gg\to h/H/A$, which are mediated
by top and bottom quark loops as well as stop and sbottom loops in the
MSSM (see Fig.~\ref{fg:lodiahgg}). Due to the large size of the top Yukawa
couplings gluon fusion comprises the dominant Higgs boson production
mechanism for small and moderate $\tgb$. For large $\tgb$ the leading
role is taken over by Higgs radiation off bottom quarks due to the
strongly enhanced bottom Yukawa couplings \cite{cxn}.

The QCD corrections to the top and bottom quark loops are known since a
long time including the full Higgs and quark mass dependences
\cite{gghnlo}. They increase the cross sections by up to about 100\%.
The limit of very heavy top quarks provides an approximation within
$\sim 20-30\%$ for $\tgb\lsim 5$ \cite{limit}. In this limit the
next-to-leading order (NLO) QCD corrections have been calculated before
\cite{gghnlolim} and later the next-to-next-to-leading order (NNLO) QCD
corrections \cite{gghnnlo}. The NNLO corrections lead to a further
moderate increase of the cross section by $\sim 20-30\%$, so that the
dominant part are the NLO contributions. Very recently an estimate of
the next-to-next-to-next-to-leading order (NNNLO) effects has been
obtained \cite{gghn3lo} indicating improved perturbative convergence.  A
full massive NNLO calculation is not available so far, so that the NNLO
results can only be used for small and moderate $\tgb$. On the other
hand the NLO corrections to the squark loops are only known in the limit
of heavy squarks \cite{gghnlosq}. Moreover, the full SUSY--QCD
corrections have been obtained recently for heavy SUSY particle masses
\cite{gghnlosqcd}.  NLO computations for the last two contributions
including the full mass dependences are missing so far. This work
presents the pure QCD corrections to the squark loops including the full
squark and Higgs mass dependences as a first step towards a full NLO
SUSY--QCD calculation.

The reverse processes, gluonic Higgs decays, play a role for the decay
profiles of the Higgs particles. Their partial widths can be measured at
a future linear $e^+e^-$ collider with energy up to about 1 TeV
\cite{tesla}. The
gluonic branching ratios can reach a level of ${\cal O}(10\%)$ in some
regions of the MSSM parameter space \cite{schladming,cxn,hdecay}.
Analogously to the gluon fusion processes the QCD corrections to the top
and bottom quark loops have been calculated including the full quark and
Higgs mass dependences \cite{gghnlo}. The NLO corrections in the limit
of very heavy top quarks can be found in \cite{gghnlolim,hggnlolim}.
This limit provides a reasonable approximation for small and moderate
values of $\tgb$. The NNLO QCD corrections are known for very heavy top
quarks, too \cite{hggnnlo}.  Very recently, even the NNNLO QCD
corrections have been obtained in the heavy top mass limit
\cite{hggnnnlo}.  While the NLO corrections enhance the partial decay
width by up to about 70\%, the NNLO corrections turn out to be of
moderate size ${\cal O}(20\%)$ supplemented by the NNNLO corrections in
the per cent range.  The quark mass effects at NNLO and beyond are
unknown. The SUSY--QCD corrections to the gluonic decay mode are
identical to the gluon fusion process. The full mass dependences of the
NLO SUSY--QCD corrections are unknown. This work provides a first
step towards this aim by determining the pure QCD corrections to the
squark loops.

A subsample of the diagrams describing the Higgs couplings to gluons
determines the Higgs couplings to a pair of photons. Since at NLO there
is no gluon radiation, the LO and NLO results for the photonic Higgs
couplings are valid for the Higgs decays into photons, which are
important Higgs decay modes for the Higgs search at the LHC. Moreover,
they determine Higgs
boson production at a future linear photon collider, a satellite mode of
a linear $e^+e^-$ collider built up by Compton back-scattered laser
light off the $e^-$ beams \cite{compton,gagatdr}. The NLO QCD corrections
to the quark loops are known in the limit of a very heavy top quark
\cite{hgaganlo,hgagalim} as well as in the fully massive case
\cite{hgaganlo}, while the NNLO \cite{hgagannlo} and NLO SUSY--QCD
\cite{hgagasqcd} corrections are only known for very heavy top and
SUSY-particle masses.  The results presented in this paper comprise a
first step towards a full massive SUSY--QCD calculation at NLO by means
of the QCD corrections to the squark loops.

This paper is organized as follows. In Section 2 we describe the NLO QCD
calculation of the squark loop contributions to the photonic and gluonic
MSSM Higgs couplings and present the results.  Section 3 summarizes and
concludes.

\section{NLO QCD Corrections}
In order to compute the pure QCD corrections to the squark loops we need
a modification of the MSSM interaction Lagrangian allowing to separate
gluon and gluino exchange contributions in a renormalizeable way. This
is achieved by starting from the basic Lagrangian\footnote{Quartic
selfinteractions among the squarks are not taken into account in this work.}
\begin{eqnarray}
{\cal L} & = & -\frac{1}{4} G^{a\mu\nu} G^a_{\mu\nu} -\frac{1}{4}
F^{\mu\nu} F_{\mu\nu}
+ \frac{1}{2}\left[ (\partial_\mu {\cal H})^2 -
M_{\cal H}^2 {\cal H}^2 \right] \\ & + & \sum_Q \left[ \bar Q(i\!\! \not
\!\! D - m_Q) Q - g_Q^{\cal H} \frac{m_Q}{v} \bar QQ {\cal H} \right] +
\sum_{\tilde Q} \left[ |D_\mu \tilde Q|^2 - m_{\tilde Q}^2 |\tilde Q|^2
- g_{\tilde Q}^{\cal H} \frac{m_{\tilde Q}^2}{v} |\tilde Q|^2 {\cal H}
\right]
\nonumber
\end{eqnarray}
with the covariant derivative $D_\mu = \partial_\mu + ig_s G^a_\mu T^a +
ie A_\mu {\cal Q}$.  Here $G^a_{\mu\nu}$ denotes the gluon field
strength tensor and $G^a_\mu$ the gluon field accompanied by the color
$SU(3)$ generators $T^a$ $(a=1,\ldots,8)$, while $F_{\mu\nu}$ is the
photon field strength tensor and $A_\mu$ the photon field associated by
the electric charge operator ${\cal Q}$. The Higgs field ${\cal H}$
represents generically either the light scalar $h$ or the heavy scalar
$H$ Higgs boson of the MSSM\footnote{Since there are no squark loop
contributions to the pseudoscalar Higgs boson couplings to photons and
gluons at leading order (LO), in this paper we will only deal with the
scalar Higgs
bosons $h,H$.}.  In this work the couplings $g_Q^{\cal H}$ and $g^{\cal
H}_{\tilde Q}$ are not renormalized, thus leading to a renormalizeable
model with strongly interacting scalars $\tilde Q$. Only if gluino
exchange contributions were taken into account, the full MSSM renormalization
would have to be performed. It should be noted that the
gluino contributions can be separated from the QCD corrections induced
by light particles due to the gluino mass.  Gluino corrections
are expected to be small \cite{gghnlosqcd}.

The numerical results in this work will be presented for the gluophobic
Higgs scenario \cite{higgsscen}, which develops strong interference
effects between quarks and squarks. It is defined by the following
choices of the MSSM parameters [$m_t = 174.3$ GeV],
\begin{eqnarray}
M_{SUSY} & = & 350~{\rm GeV} \non \\
\mu=M_2 & = & 300~{\rm GeV} \non \\
X_t & = & -770~{\rm GeV} \non \\
A_b & = & A_t \non \\
m_{\tilde g} & = & 500~{\rm GeV} \, ,
\end{eqnarray}
where $X_t = A_t - \mu/\tgb$.
We have used the program HDECAY \cite{hdecay} for the numerical
determination of the SUSY particle masses and couplings. In this
scenario the squark masses amount to \\[-0.9cm]
\begin{eqnarray}
\tgb=3:  \qquad m_{\tilde t_1} & = & 156~{\rm GeV} \non \\
                m_{\tilde t_2} & = & 517~{\rm GeV} \non \\
                m_{\tilde b_1} & = & 346~{\rm GeV} \non \\
                m_{\tilde b_2} & = & 358~{\rm GeV} \non \\[0.1cm]
\tgb=30: \qquad m_{\tilde t_1} & = & 155~{\rm GeV} \non \\
                m_{\tilde t_2} & = & 516~{\rm GeV} \non \\
                m_{\tilde b_1} & = & 314~{\rm GeV} \non \\
                m_{\tilde b_2} & = & 388~{\rm GeV} \, .
\end{eqnarray}
The gluophobic scenario maximizes the destructive interference effects
between top and stop loop contributions to the light scalar Higgs
coupling to gluons. The results of this work look similar, whenever the
squark masses turn out to be of the order of the top mass, or the Higgs
mass reaches values beyond the corresponding squark-antisquark
threshold.

\subsection{Scalar Higgs Boson Couplings to Photons}
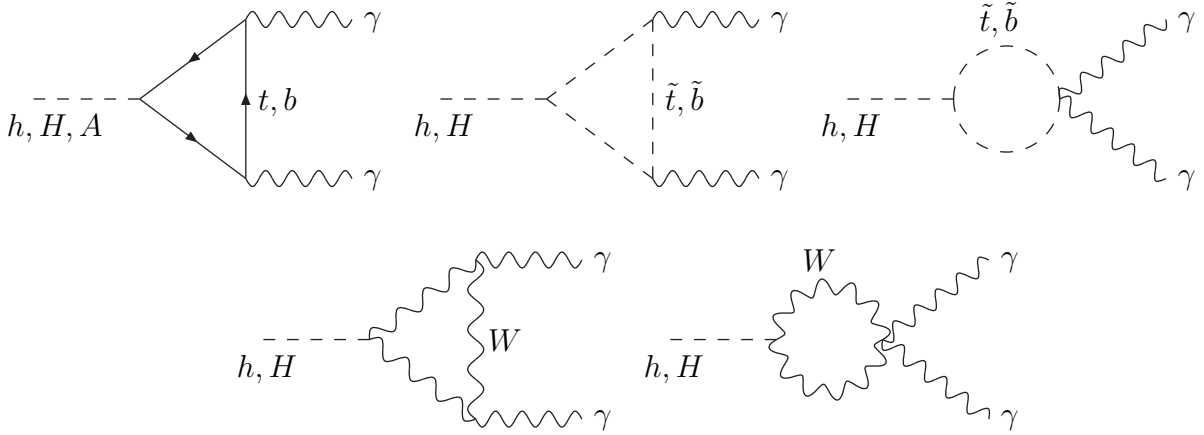
\begin{figure}[htb]
\begin{center}
\begin{picture}(130,90)(20,0)
\Photon(90,20)(130,20){-3}{4}
\Photon(90,80)(130,80){3}{4}
\ArrowLine(90,20)(90,80)
\ArrowLine(90,80)(50,50)
\ArrowLine(50,50)(90,20)
\DashLine(10,50)(50,50){5}
\put(135,18){$\gamma$}
\put(135,78){$\gamma$}
\put(95,46){$t,b$}
\put(0,36){$h,H,A$}
\end{picture}
\begin{picture}(130,90)(0,0)
\Photon(90,20)(130,20){-3}{4}
\Photon(90,80)(130,80){3}{4}
\DashLine(90,20)(90,80){5}
\DashLine(90,80)(50,50){5}
\DashLine(50,50)(90,20){5}
\DashLine(10,50)(50,50){5}
\put(135,18){$\gamma$}
\put(135,78){$\gamma$}
\put(95,46){$\tilde t,\tilde b$}
\put(0,36){$h,H$}
\end{picture}
\begin{picture}(130,90)(-20,0)
\Photon(90,50)(130,20){3}{5}
\Photon(90,50)(130,80){3}{5}
\DashCArc(70,50)(20,180,360){5}
\DashCArc(70,50)(20,0,180){5}
\DashLine(10,50)(50,50){5}
\put(135,18){$\gamma$}
\put(135,78){$\gamma$}
\put(60,76){$\tilde t,\tilde b$}
\put(0,36){$h,H$}
\end{picture}  \\
\begin{picture}(130,90)(0,0)
\Photon(90,20)(130,20){-3}{4}
\Photon(90,80)(130,80){3}{4}
\Photon(90,20)(90,80){3}{4}
\Photon(90,80)(50,50){3}{4}
\Photon(50,50)(90,20){3}{4}
\DashLine(10,50)(50,50){5}
\put(135,18){$\gamma$}
\put(135,78){$\gamma$}
\put(95,46){$W$}
\put(0,36){$h,H$}
\end{picture}
\begin{picture}(130,90)(-20,0)
\Photon(90,50)(130,20){-3}{5}
\Photon(90,50)(130,80){-3}{5}
\PhotonArc(70,50)(20,180,360){3}{6}
\PhotonArc(70,50)(20,0,180){3}{6}
\DashLine(10,50)(50,50){5}
\put(135,18){$\gamma$}
\put(135,78){$\gamma$}
\put(60,76){$W$}
\put(0,36){$h,H$}
\end{picture}  \\
\caption{\label{fg:lodiahgaga} \it MSSM Higgs boson couplings to photon
pairs mediated by top- and bottom quark, stop and sbottom as well as $W$
boson loops at leading order.}
\end{center}
\end{figure}

At leading order (LO) the photonic MSSM Higgs couplings are mediated by
top and bottom quark loops as well as $W$ loops for the scalar Higgs
particles $h,H$.  If the stop and sbottom masses range below $\sim 400$
GeV, there are significant contributions from squark loops, too (see
Fig.~\ref{fg:lodiahgaga}).
The LO photonic decay widths are given by \cite{cxn,gghnlo,ellis}
\begin{eqnarray}
\Gamma_{LO}(h/H\to \gamma\gamma) & = &
\frac{G_F\alpha^2M_{h/H}^3}{36\sqrt{2}\pi^3} \left|g^{h/H}_W
A_W^{h/H}(\tau_W) + \sum_f N_{cf} e_f^2 g_f^{h/H} A_f^{h/H}(\tau_f)
\right.  \label{eq:hgagalo} \\
& & \qquad \qquad \qquad \left. + \sum_{\tilde f} N_{c\tilde f}
e_{\tilde f}^2 g_{\tilde f}^{h/H}
A_{\tilde f}^{h/H} (\tau_{\tilde f}) \right|^2 \nonumber \\
A_W^{h/H}(\tau) & = & - [2+3\tau+3\tau(2-\tau)f(\tau)] \nonumber \\
A_f^{h/H}(\tau) & = & 2 \tau [1+(1-\tau)f(\tau)] \nonumber \\
A_{\tilde f}^{h/H} (\tau) & = & -\tau[1-\tau f(\tau)]
\nonumber \\
f(\tau) & = & \left\{ \begin{array}{ll}
\displaystyle \arcsin^2 \frac{1}{\sqrt{\tau}} & \tau \ge 1 \\
\displaystyle - \frac{1}{4} \left[ \log \frac{1+\sqrt{1-\tau}}
{1-\sqrt{1-\tau}} - i\pi \right]^2 & \tau < 1
\end{array} \right. \nonumber \, ,
\end{eqnarray}
where we neglected contributions from charginos, charged Higgs bosons
and the charged sleptons. The full expressions can be found e.g.\ in
\cite{cxn} and have been used in our work. $N_{cf} (N_{c\tilde f})$
denote the color factors and $e_f (e_{\tilde f})$ the electric charges
of the (s)fermions in units of the proton charge, while the scaling
variables are defined as $\tau_i = 4m_i^2/M_{h/H}^2$.  For large loop
particle masses the form factors approach constant values,
\begin{eqnarray*}
A_f^{h/H} (\tau) & \to & \frac{4}{3} \hspace*{1cm}
\mbox{for $M_{h/H}^2 \ll 4m_f^2$} \nonumber \\
A_{\widetilde{f}}^{h/H} (\tau) & \to & \frac{1}{3} \hspace*{1cm}
\mbox{for $M_{h/H}^2 \ll 4m_{\widetilde{f}}^2$} \nonumber \\
A_W^{h/H} (\tau) & \to & - 7 \hspace*{1cm} \mbox{for $M_{h/H}^2 \ll
4M_W^2$}
\, .
\end{eqnarray*}

The reverse processes, $\gamma\gamma \to h/H$, play an important role
for the MSSM Higgs search at a photon collider at high energies
\cite{gagatdr,gagah}. Such a photon collider can be realized by Compton
back-scattering of laser light from high-energy electron beams in a
linear $e^+e^-$ collider \cite{compton}. In this collider $\gamma\gamma$
c.m.\ energies up to about 80\% of the corresponding $e^+e^-$ c.m.\
energy can be reached.  The $s$-channel Higgs boson production cross
sections are then given by
\begin{equation}
\langle \sigma(\gamma\gamma\to h/H) \rangle = \frac{8\pi^2}{M_{h/H}^3}
\Gamma(h/H\to\gamma\gamma) \frac{d{\cal L}^{\gamma\gamma}}{d\tau_{h/H}}
\, ,
\end{equation}
where $d{\cal L}^{\gamma\gamma}/d\tau_{h/H}$ denotes the differential
$\gamma\gamma$ luminosity for $\tau_{h/H} = M_{h/H}^2/s_{\gamma\gamma}$
with $s_{\gamma\gamma}$ being the squared c.m.~energy of the photon collider.
This relation between the photon fusion cross section and the photonic
Higgs decay width holds in NLO QCD, too, since single gluon radiation
vanishes due to the conservation of color charges as well as due to the
Furry theorem.

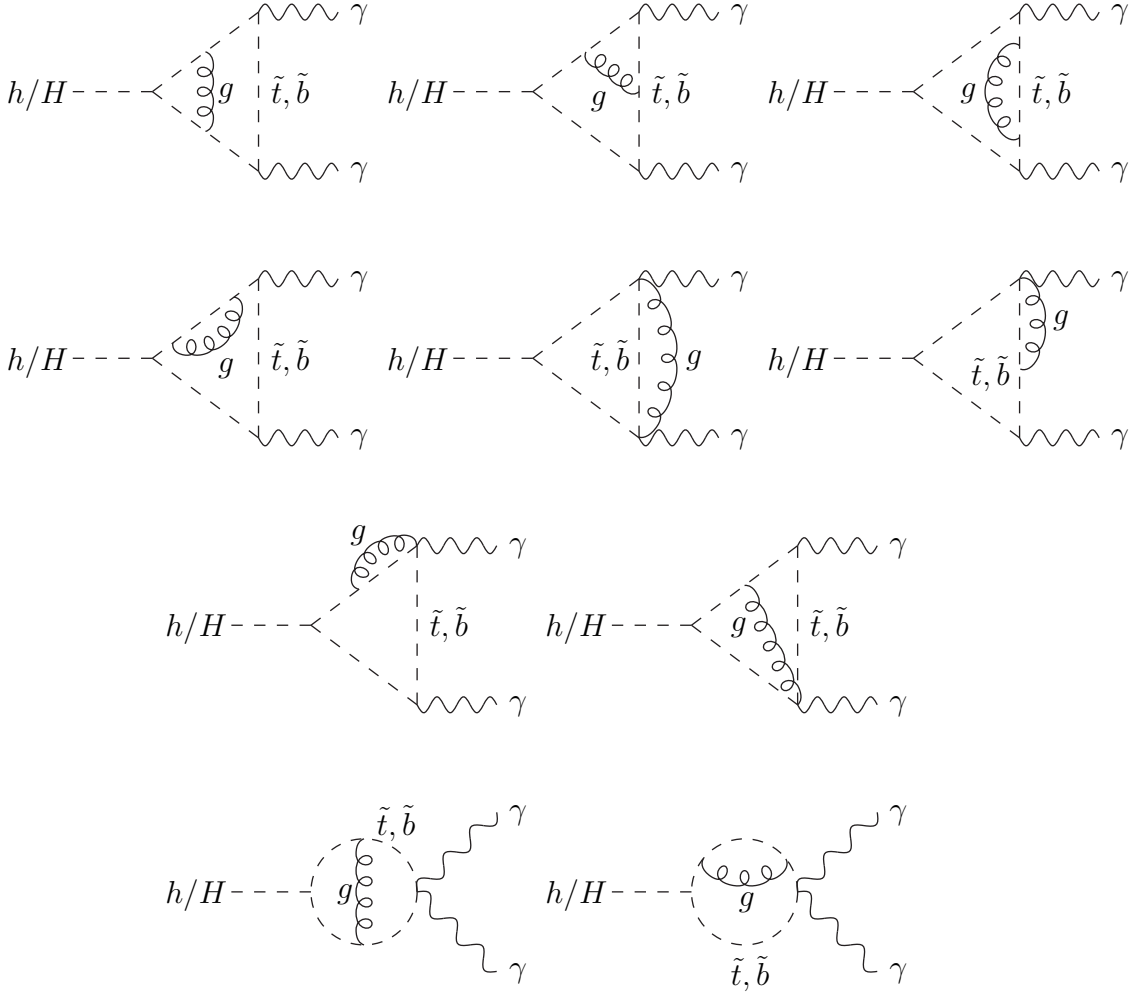
\begin{figure}[hbt]
\begin{picture}(100,90)(-40,0)
\Photon(70,20)(100,20){-3}{3}
\Photon(70,80)(100,80){3}{3}
\Gluon(50,35)(50,65){-3}{3}
\DashLine(70,20)(70,80){5}
\DashLine(70,80)(30,50){5}
\DashLine(30,50)(70,20){5}
\DashLine(0,50)(30,50){5}
\put(-25,46){$h/H$}
\put(75,46){$\tilde t,\tilde b$}
\put(105,18){$\gamma$}
\put(105,78){$\gamma$}
\put(55,48){$g$}
\end{picture}
\begin{picture}(100,90)(-80,0)
\Photon(70,20)(100,20){-3}{3}
\Photon(70,80)(100,80){3}{3}
\Gluon(70,50)(50,65){3}{3}
\DashLine(70,20)(70,80){5}
\DashLine(70,80)(30,50){5}
\DashLine(30,50)(70,20){5}
\DashLine(0,50)(30,50){5}
\put(-25,46){$h/H$}
\put(75,46){$\tilde t,\tilde b$}
\put(105,18){$\gamma$}
\put(105,78){$\gamma$}
\put(52,45){$g$}
\end{picture}
\begin{picture}(100,90)(-120,0)
\Photon(70,20)(100,20){-3}{3}
\Photon(70,80)(100,80){3}{3}
\GlueArc(80,50)(20,120,240){3}{4}
\DashLine(70,20)(70,80){5}
\DashLine(70,80)(30,50){5}
\DashLine(30,50)(70,20){5}
\DashLine(0,50)(30,50){5}
\put(-25,46){$h/H$}
\put(75,46){$\tilde t,\tilde b$}
\put(105,18){$\gamma$}
\put(105,78){$\gamma$}
\put(48,48){$g$}
\end{picture} \\
\begin{picture}(100,100)(-40,0)
\Photon(70,20)(100,20){-3}{3}
\Photon(70,80)(100,80){3}{3}
\GlueArc(45,70)(16,243,371){3}{4}
\DashLine(70,20)(70,80){5}
\DashLine(70,80)(30,50){5}
\DashLine(30,50)(70,20){5}
\DashLine(0,50)(30,50){5}
\put(-25,46){$h/H$}
\put(75,46){$\tilde t,\tilde b$}
\put(105,18){$\gamma$}
\put(105,78){$\gamma$}
\put(55,45){$g$}
\end{picture}
\begin{picture}(100,100)(-80,0)
\Photon(70,20)(100,20){-3}{3}
\Photon(70,80)(100,80){3}{3}
\GlueArc(38,50)(43.5,-43,43){3}{5}
\DashLine(70,20)(70,80){5}
\DashLine(70,80)(30,50){5}
\DashLine(30,50)(70,20){5}
\DashLine(0,50)(30,50){5}
\put(-25,46){$h/H$}
\put(52,46){$\tilde t,\tilde b$}
\put(105,18){$\gamma$}
\put(105,78){$\gamma$}
\put(88,48){$g$}
\end{picture}
\begin{picture}(100,100)(-120,0)
\Photon(70,20)(100,20){-3}{3}
\Photon(70,80)(100,80){3}{3}
\GlueArc(52,63)(25,-43,43){3}{3}
\DashLine(70,20)(70,80){5}
\DashLine(70,80)(30,50){5}
\DashLine(30,50)(70,20){5}
\DashLine(0,50)(30,50){5}
\put(-25,46){$h/H$}
\put(52,40){$\tilde t,\tilde b$}
\put(105,18){$\gamma$}
\put(105,78){$\gamma$}
\put(83,63){$g$}
\end{picture} \\
\begin{picture}(100,100)(-100,0)
\Photon(70,20)(100,20){-3}{3}
\Photon(70,80)(100,80){3}{3}
\GlueArc(64,65)(16,69,185){3}{4}
\DashLine(70,20)(70,80){5}
\DashLine(70,80)(30,50){5}
\DashLine(30,50)(70,20){5}
\DashLine(0,50)(30,50){5}
\put(-25,46){$h/H$}
\put(75,46){$\tilde t,\tilde b$}
\put(105,18){$\gamma$}
\put(105,78){$\gamma$}
\put(45,83){$g$}
\end{picture}
\begin{picture}(100,100)(-140,0)
\Photon(70,20)(100,20){-3}{3}
\Photon(70,80)(100,80){3}{3}
\Gluon(50,65)(70,20){3}{5}
\DashLine(70,20)(70,80){5}
\DashLine(70,80)(30,50){5}
\DashLine(30,50)(70,20){5}
\DashLine(0,50)(30,50){5}
\put(-25,46){$h/H$}
\put(75,46){$\tilde t,\tilde b$}
\put(105,18){$\gamma$}
\put(105,78){$\gamma$}
\put(45,48){$g$}
\end{picture} \\
\begin{picture}(100,100)(-100,0)
\Photon(70,50)(100,20){3}{3}
\Photon(70,50)(100,80){3}{3}
\Gluon(50,70)(50,30){-3}{4}
\DashCArc(50,50)(20,0,360){4}
\DashLine(0,50)(30,50){5}
\put(-25,46){$h/H$}
\put(55,72){$\tilde t,\tilde b$}
\put(105,18){$\gamma$}
\put(105,78){$\gamma$}
\put(40,48){$g$}
\end{picture}
\begin{picture}(100,100)(-140,0)
\Photon(70,50)(100,20){3}{3}
\Photon(70,50)(100,80){3}{3}
\GlueArc(50,75)(20,219,321){3}{3}
\DashCArc(50,50)(20,0,360){4}
\DashLine(0,50)(30,50){5}
\put(-25,46){$h/H$}
\put(45,15){$\tilde t,\tilde b$}
\put(105,18){$\gamma$}
\put(105,78){$\gamma$}
\put(48,45){$g$}
\end{picture} \\[-1cm]
\caption[]{\label{fg:hgagadianlo} \it Generic diagrams for the NLO QCD
corrections to the squark contributions to the photonic Higgs
couplings.}
\end{figure}
The NLO QCD corrections to the photonic Higgs decay modes and the photon
fusion cross section require the calculation of the two-loop diagrams
depicted in Fig.~\ref{fg:hgagadianlo}. We have reduced the 5-dimensional
Feynman parameter integrals to one-dimensional ones which have been
integrated numerically\footnote{Analytical results can be derived, too,
as in the case of the quark loops \cite{harlander}.}. In a second
calculation the QCD corrections have been obtained purely numerically.
Both results agree within integration errors. The QCD corrections can be
parametrized by shifts of the quark and squark amplitudes according to
\begin{eqnarray}
A_Q^{h/H} (\tau_Q) & \to & A_Q^{h/H} (\tau_Q) \left[ 1+C_Q^{h/H}(\tau_Q)
\frac{\alpha_s}{\pi} \right] \label{eq:gagaamp} \\
A_{\tilde Q}^{h/H} (\tau_{\tilde Q}) & \to & A_{\tilde Q}^{h/H}
(\tau_{\tilde Q}) \left[ 1+C_{\tilde Q}^{h/H}(\tau_{\tilde Q})
\frac{\alpha_s}{\pi} \right] \, .
\nonumber
\end{eqnarray}

\begin{figure}[hbt]
\vspace*{-2.0cm}
\hspace*{2.0cm}
\epsfxsize=12cm \epsfbox{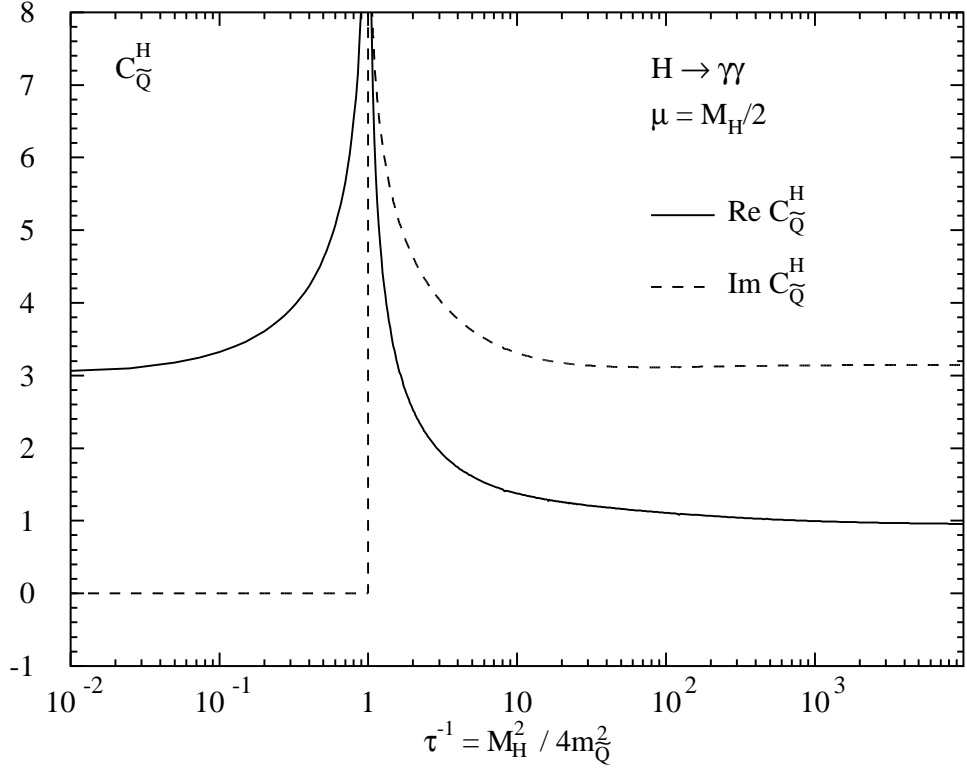}
\vspace*{-4.0cm}
\caption[]{\label{fg:csqgaga} \it Real and imaginary parts of the QCD
correction factor to the squark amplitudes of the two-photon couplings
for the scalar MSSM Higgs bosons. The renormalization scale of the
running squark mass is taken to be $\mu=M_{\cal H}/2$.}
\end{figure}

The QCD corrections to the quark loops can be found in
Refs.~\cite{hgaganlo}. The corresponding coefficient $C_{\tilde
Q}^{h/H}(\tau_{\tilde Q})$ of the QCD corrections to the squark loops
for finite Higgs and squark masses is presented in
Fig.~\ref{fg:csqgaga} as a function of $\tau_{\tilde Q}$. In order to
improve the perturbative behavior of the squark loop contributions they
should be expressed preferably in terms of the running squark masses
$m_{\tilde Q}(M_{\cal H}/2)$, which are related to the pole
masses $M_{\tilde Q}$ via
\begin{equation}
m_{\tilde Q}(\mu) = M_{\tilde Q} \left(
\frac{\alpha_s(\mu)}{\alpha_s(M_{\tilde Q})}\right)^{\frac{6}{\beta_0}}
\end{equation}
where $\beta_0 = 33-2N_F$ with $N_F=5$ light flavors.
Their scale is identified with $\mu=M_{\cal H}/2$ within the photonic
decay mode. These definitions imply a proper definition of the $\tilde
Q \bar{\tilde Q}$ thresholds $M_{\cal H} = 2 M_{\tilde Q}$, without
artificial displacements due to finite shifts between the {\it pole} and
running squark masses, as is the case for the running $\overline{\rm MS}$
masses. We have taken into account the LO scale dependence of the squark
masses generated by light particle contributions.

The coefficient $C_{\tilde Q}^{h/H}(\tau_{\tilde Q})$ is real below the
$\tilde Q \bar{\tilde Q}$ threshold and complex above. For
$\tau_{\tilde Q}=1$ it diverges, so that within a margin of a few GeV
around the threshold the
perturbative analysis cannot be applied. This singular behavior
originates from a Coulomb singularity at the threshold, since $\tilde
Q\bar{\tilde Q}$ pairs can form $0^{++}$ states. This
implies that the imaginary part $\Im m~C_{\tilde Q}^{h/H}$ develops a step
due to Coulombic gluon exchange, thus resulting in a singular behavior
of the real part at threshold.
The singular behavior can be quantified. The lowest order form factor
of the squark loops is given by
\begin{equation}
A_{\tilde Q}^{h/H, LO} (\tau) = -\tau[1-\tau f(\tau)] \to
\frac{\pi^2}{4} - 1 + i\pi\beta \qquad \mbox
{for}~\tau \to 1
\end{equation}
where $\beta=\sqrt{1-\tau}$ denotes the squark velocity above threshold.
The QCD corrections to the imaginary part at threshold can be derived
from the Sommerfeld rescattering correction \cite{sommerfeld},
\begin{equation}
C_{Coul} = \frac{Z}{1-e^{-Z}} = 1 + \frac{Z}{2} + {\cal O}(\alpha_s^2)
\qquad \mbox{for}~ Z = \frac{4\pi\alpha_s}{3\beta} \, .
\end{equation}
The QCD corrected imaginary part of the $h/H\gamma\gamma$ couplings at
threshold is now given as
\begin{equation}
\Im m~A_{\tilde Q}^{h/H} = \pi\beta C_{coul} = \pi\beta + \frac{2\pi^2}{3}
\alpha_s \, ,
\end{equation}
so that the real part can be derived from a once-subtracted dispersion
relation yielding the following expression in the threshold region
\begin{equation}
A_{\tilde Q}^{h/H} \to A_{\tilde Q}^{h/H, LO} + \frac{2\pi\alpha_s}{3}
\left[ -\log(\tau_{\tilde Q}^{-1}-1) + i\pi + const \right] \, ,
\end{equation}
where the smooth non-singular constant is not relevant for the divergent
behavior. Finally the singular behavior of the QCD correction factor
at threshold can be cast into the form
\begin{eqnarray}
\Re e~C_{\tilde Q}^{h/H} & \to &
-\frac{8\pi^2}{3\left(\pi^2-4\right)} \log(\tau_{\tilde
Q}^{-1}-1) + const \nonumber \\
\Im m~C_{\tilde Q}^{h/H} & \to &
\frac{8\pi^3}{3\left(\pi^2-4\right)} \approx 14.09 \, .
\label{eq:coul}
\end{eqnarray}
The total size of the imaginary part and the singular behavior of the
real part agree with the numerical results depicted in
Fig.~\ref{fg:csqgaga}. However, the singular behaviour at threshold is
unphysical. A proper inclusion of the finite decay widths of the virtual
squarks as well as a resummation of the Coulomb singularities
\cite{threshold} will
regularize the divergences and improve the perturbative results at
threshold.

In the limit of heavy squark masses the coefficient $C_{\tilde
Q}^{h/H}(\tau_{\tilde Q})$ approaches a finite and constant
value\footnote{This value differs from the heavy squark limit obtained
in Ref.~\cite{hgagasqcd}. The difference can be traced back to a wrong
expression for the anomalous squark mass dimension used in \cite{hgagasqcd}.}
\begin{equation}
C_{\tilde Q}^{h/H}(\tau_{\tilde Q}) \to 3 \, .
\label{eq:csqlim}
\end{equation}
This asymptotic value can also be derived from low-energy theorems
\cite{cxn,gghnlo,gghnlosq,let}, which are based on the relation between
the matrix elements with and without a light external Higgs boson. This
allows the derivation of an effective Lagrangian for the Higgs couplings
to photons mediated by squark loops in the heavy squark mass
limit\footnote{The anomalous dimension of the kinetic photon operator
$F^{\mu\nu}F_{\mu\nu}$ does not contribute at NLO \cite{limit,hggnlolim}.}
\begin{equation}
\Delta {\cal L}_{eff} = \frac{e_{\tilde Q}^2}{4} \frac{\beta_{\tilde
Q}(\alpha)/\alpha}{1+ \gamma_{m_{\tilde Q}}(\alpha_s)} F^{\mu\nu}
F_{\mu\nu} \frac{H}{v} \, ,
\end{equation}
where $\beta_{\tilde Q}(\alpha)/\alpha = (\alpha/2\pi)
[1+4\alpha_s/\pi+\cdots]$ denotes the heavy squark $\tilde Q$
contribution to the QED $\beta$ function and $\gamma_{m_{\tilde Q}}
(\alpha_s)= \alpha_s/\pi + \cdots$ the anomalous squark mass dimension.  The
NLO expansion of the effective Lagrangian reads as
\begin{equation}
\Delta {\cal L}_{eff} = e_{\tilde Q}^2 \frac{\alpha}{8\pi} F^{\mu\nu} F_{\mu\nu}
\frac{H}{v}
\left[1 + 3 \frac{\alpha_s}{\pi} + {\cal O}(\alpha_s^2) \right] \, ,
\label{eq:leffhgaga}
\end{equation}
which agrees with the $C_{\tilde Q}^{h/H}$-value of
Eq.~(\ref{eq:csqlim}) in the heavy squark limit.

\begin{figure}[hbtp]
\begin{picture}(100,500)(0,0)
\put(40.0,120.0){\includegraphics{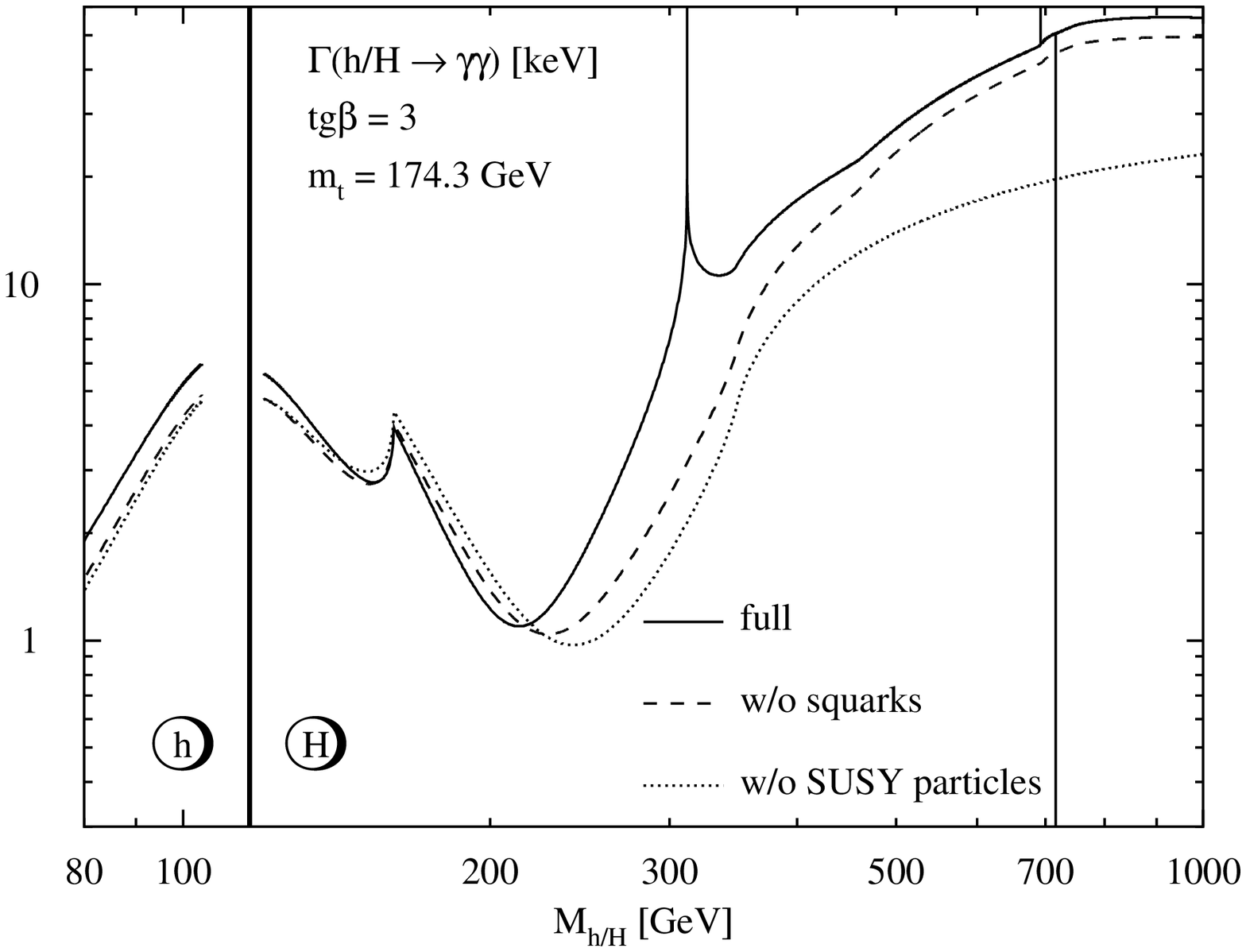}}
\put(40.0,-135.0){\includegraphics{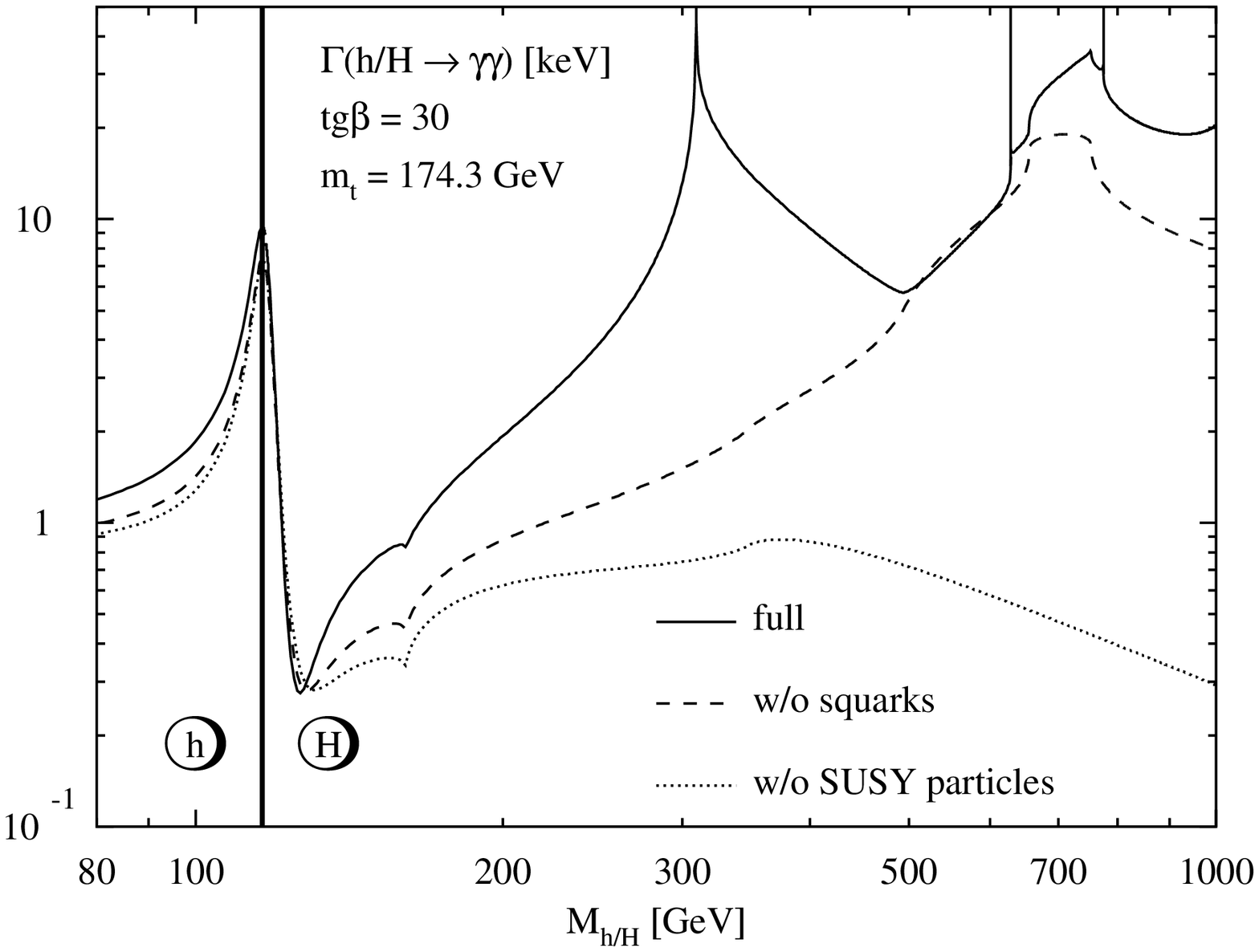}}
\end{picture}
\caption[]{\label{fg:ghgaga} \it QCD corrected partial decay widths of
the scalar MSSM Higgs bosons to two photons as functions of the
corresponding Higgs masses for $\tgb=3$ and 30. The full curves include
all loop contributions, in the dashed lines the squark contributions are
omitted and in the dotted lines all SUSY particle loops are neglected.
The kinks, bumps and spikes correspond to the $WW, \tilde t_1\bar{\tilde
t}_1, t\bar t, \tilde b_1\bar{\tilde b}_1, \tilde \tau_1 \bar{\tilde
\tau}_1, \tilde \tau_2 \bar{\tilde \tau}_2$ and $\tilde b_2\bar{\tilde
b}_2$ thresholds in consecutive order with rising Higgs mass. The
renormalization scale of the running quark and squark masses is chosen
to be $M_{h/H}/2$, while the scale of $\alpha_s$ is taken to be the
corresponding Higgs mass.}
\end{figure}

\begin{figure}[hbtp]
\begin{picture}(100,500)(0,0)
\put(40.0,120.0){\includegraphics{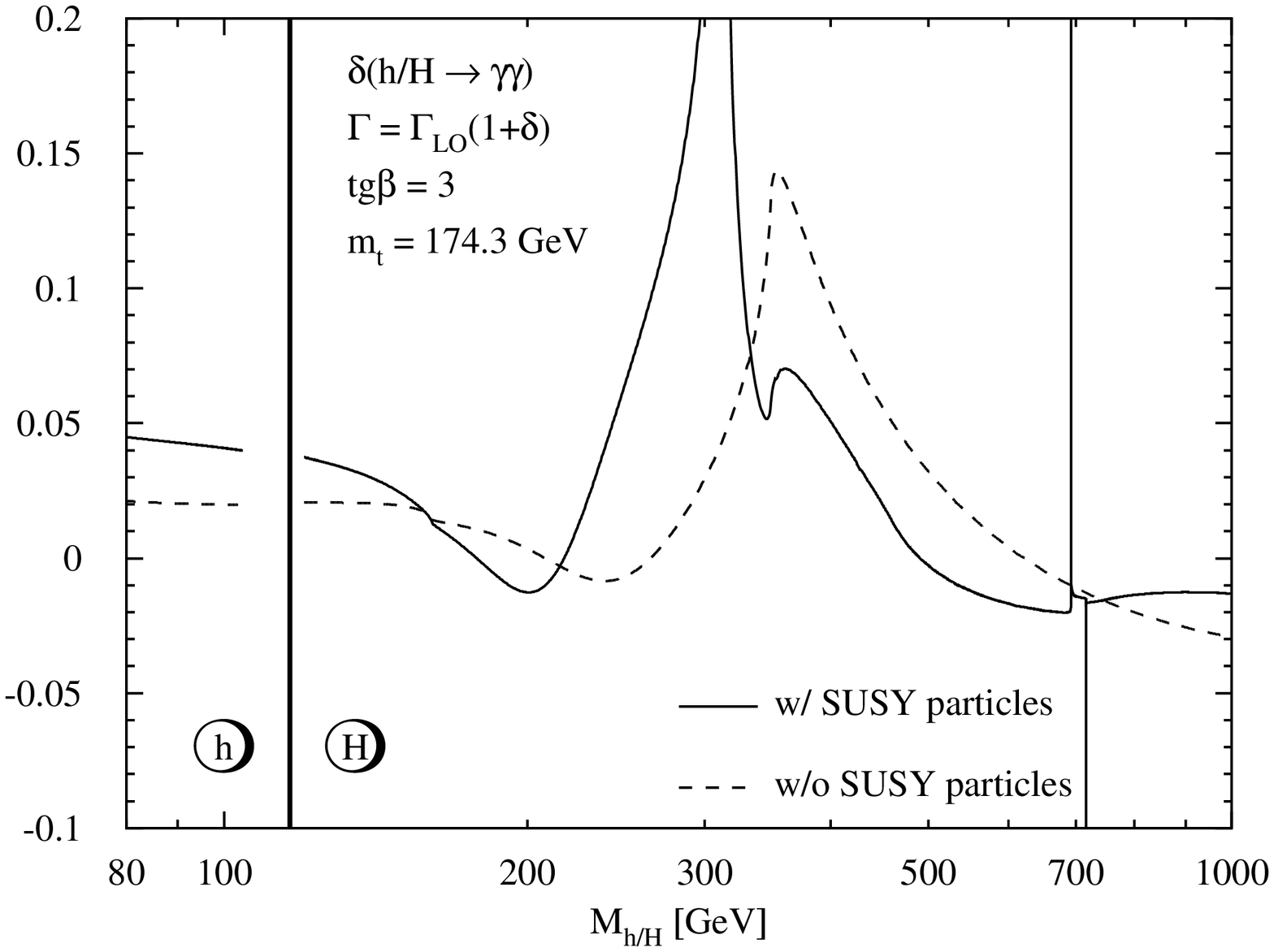}}
\put(40.0,-135.0){\includegraphics{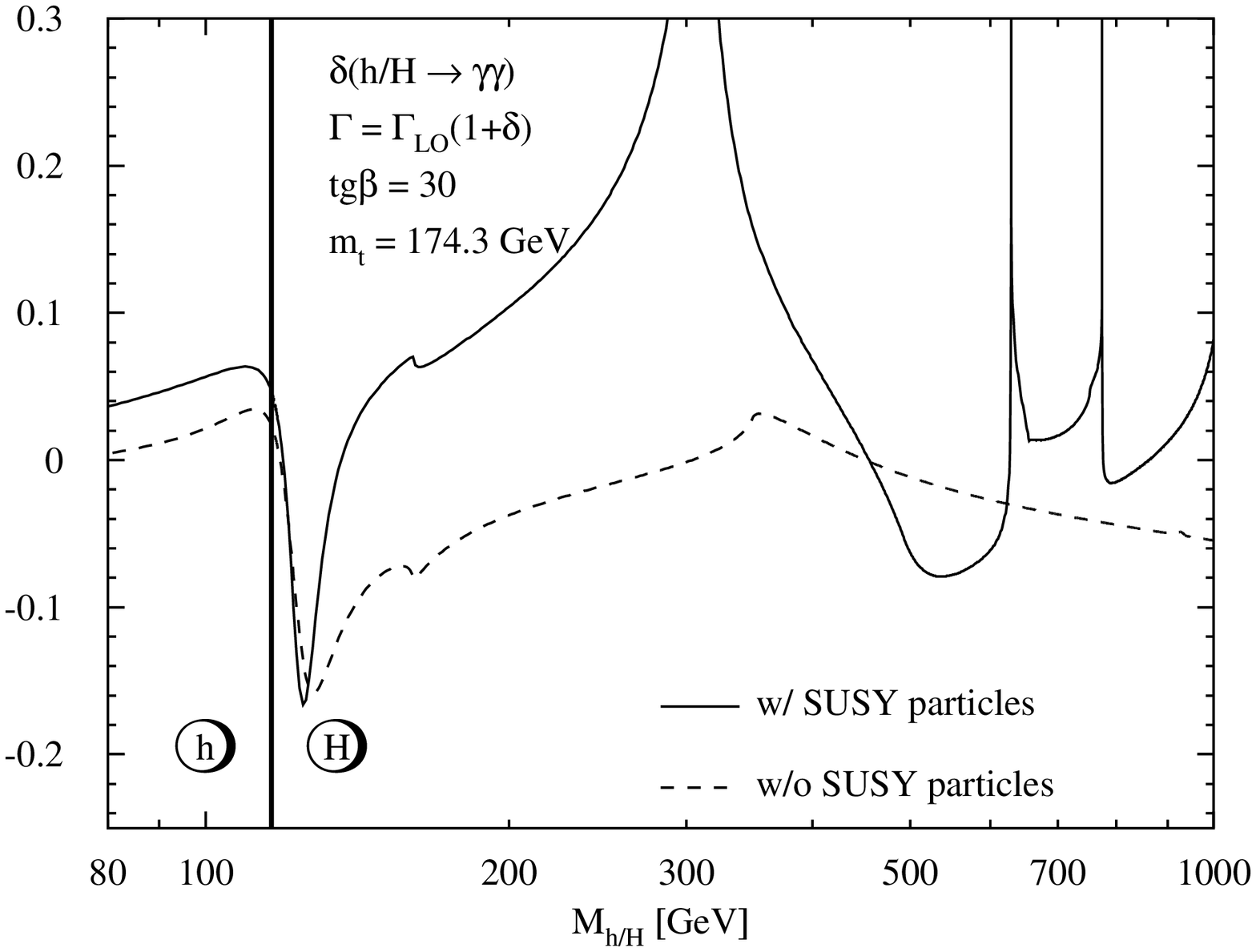}}
\end{picture}
\caption[]{\label{fg:dhgaga} \it Relative QCD corrections to the partial
decay widths of the scalar MSSM Higgs bosons to two photons as functions
of the corresponding Higgs masses for $\tgb=3$ and 30. The full curves
include all loop contributions while in the dashed lines the squark
contributions are omitted. The kinks and spikes correspond to the $WW,
\tilde t_1\bar{\tilde t}_1, t\bar t, \tilde b_1\bar{\tilde b}_1, \tilde
\tau_1 \bar{\tilde \tau}_1, \tilde \tau_2 \bar{\tilde\tau}_2$ and
$\tilde b_2\bar{\tilde b}_2$ thresholds in consecutive order with rising
Higgs mass. The renormalization scale of the running quark and squark
masses is chosen to be $M_{h/H}/2$, while the scale of $\alpha_s$ is
taken to be the corresponding Higgs mass.}
\end{figure}

\begin{figure}[hbtp]
\begin{picture}(100,500)(0,0)
\put(40.0,120.0){\includegraphics{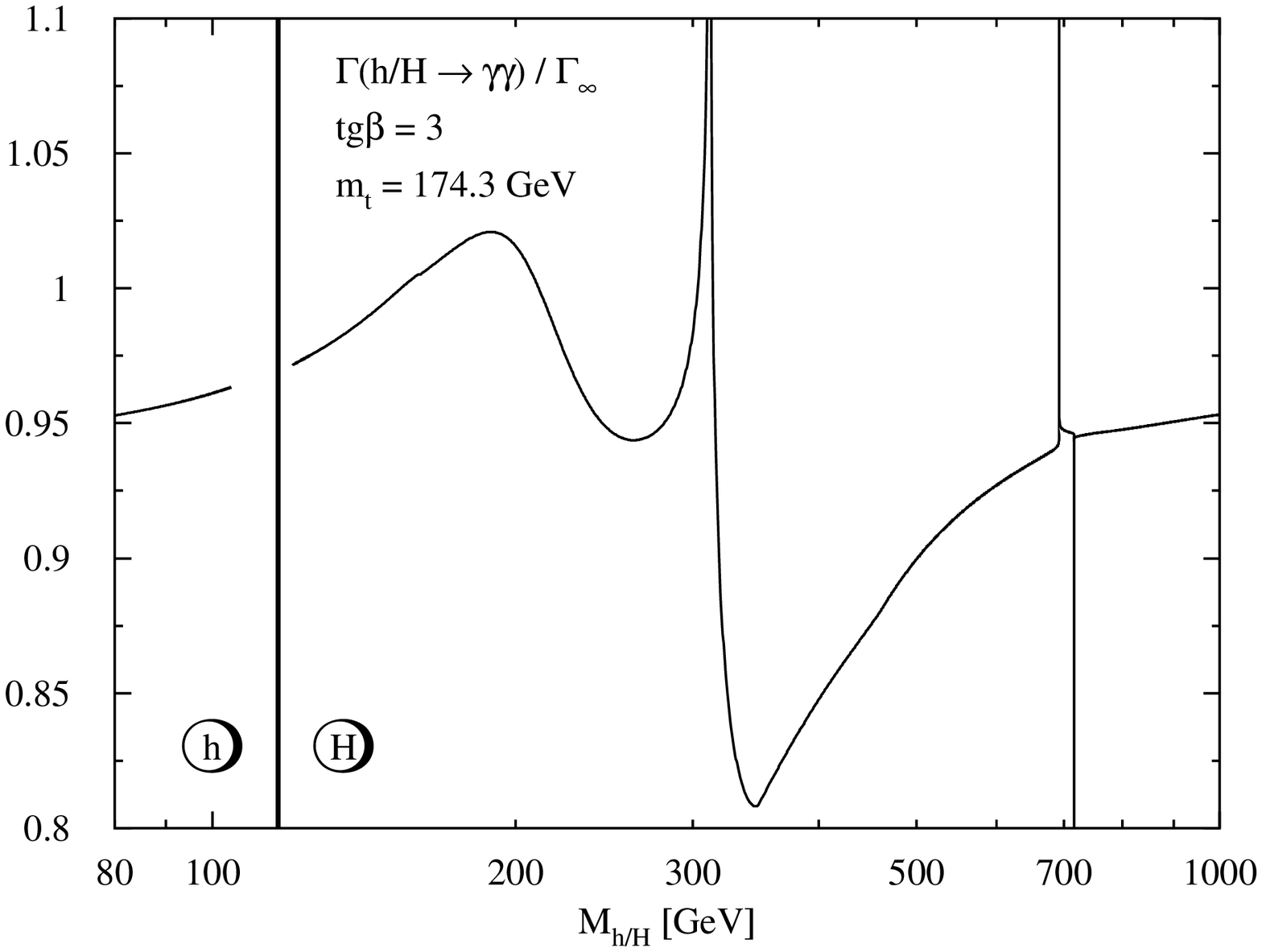}}
\put(40.0,-135.0){\includegraphics{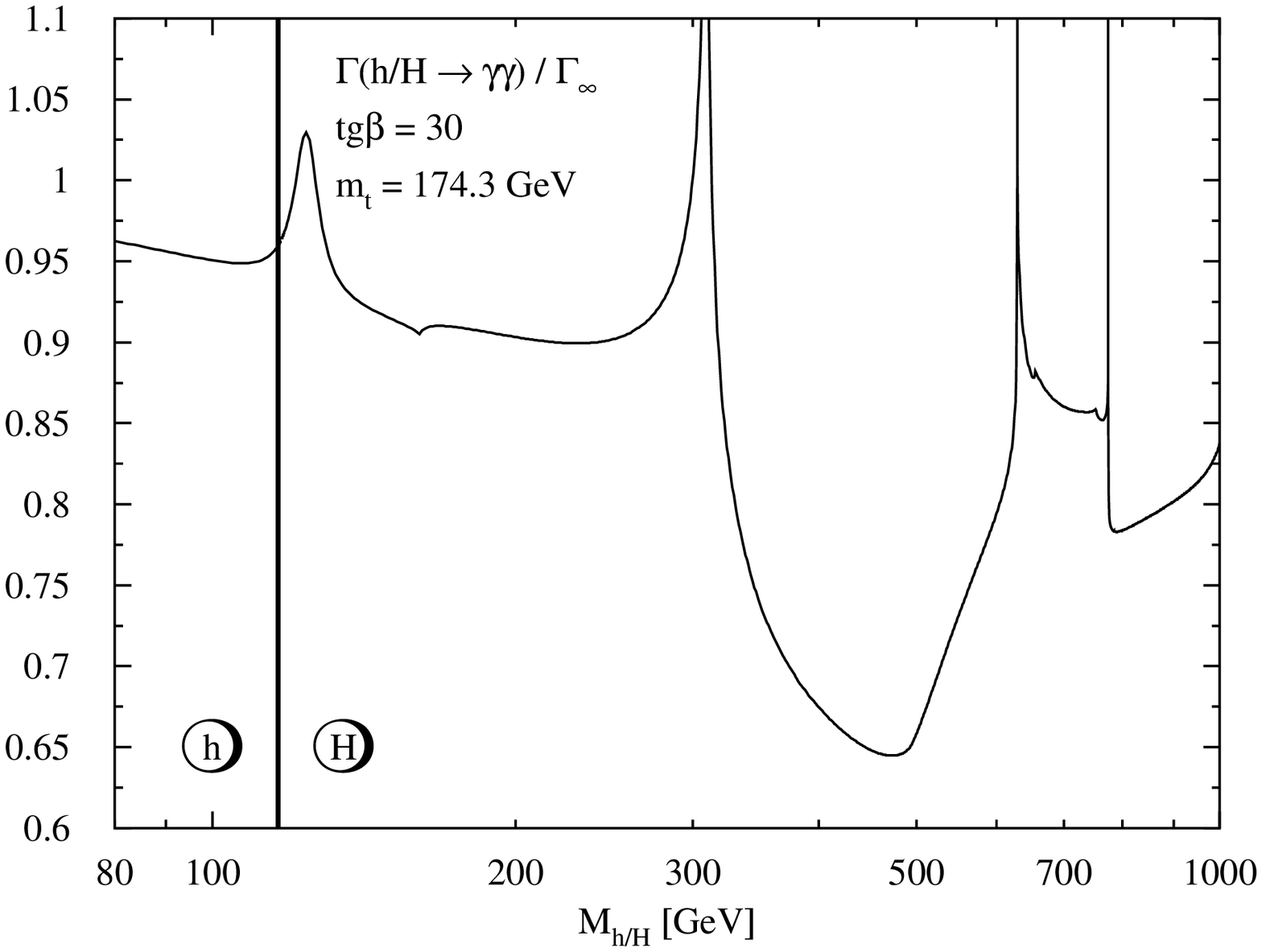}}
\end{picture}
\caption[]{\label{fg:mhgaga} \it Ratio of the QCD corrected partial
decay widths of the scalar MSSM Higgs bosons to two photons including
the full squark mass dependence and those obtained by taking the
relative QCD corrections to the squark loops in the heavy mass limit as
functions of the corresponding Higgs masses for $\tgb=3$ and 30.  The
kinks and spikes correspond to the $WW, \tilde t_1\bar{\tilde t}_1,
t\bar t, \tilde b_1\bar{\tilde b}_1, \tilde \tau_1 \bar{\tilde \tau}_1,
\tilde \tau_2 \bar{\tilde\tau}_2$ and $\tilde b_2\bar{\tilde b}_2$
thresholds in consecutive order with rising Higgs mass. The
renormalization scale of the running quark and squark masses is chosen
to be $M_{h/H}/2$, while the scale of $\alpha_s$ is taken to be the
corresponding Higgs mass.}
\end{figure}

The numerical results are presented in
Figs.~\ref{fg:ghgaga}--\ref{fg:mhgaga} for the gluophobic Higgs
scenario\footnote{The massive QCD corrections to the quark and squark
loops have been implemented in the program HDECAY \cite{hdecay}.}.
Fig.~\ref{fg:ghgaga} shows the partial photonic widths at NLO of the
scalar Higgs bosons $h,H$ for two values of $\tgb=3,30$. The dotted
lines exhibit the photonic widths without the contributions of SUSY
particles, the dashed lines include all SUSY particle contributions
except those of the squarks, while the full curves correspond to the
full partial decay widths. For small as well as for large values of
$\tgb$ the photonic widths range between about 0.1 and 10 keV. The
importance of the SUSY particles and in particular the squark
contributions is clearly visible from the comparison of the three
different curves. The kinks, bumps and spikes correspond to the $WW,
\tilde t_1\bar{\tilde t}_1, t\bar t, \tilde b_1\bar{\tilde b}_1, \tilde
\tau_1\bar{\tilde \tau}_1, \tilde \tau_2\bar{\tilde \tau}_2$ and $\tilde
b_2\bar{\tilde b}_2$ thresholds in consecutive order\footnote{In the
gluophobic scenario the $\tilde \tau$ masses amount to
$m_{\tilde\tau_1}=348$ GeV, $m_{\tilde\tau_2}=356$ GeV for $\tgb=3$ and
$m_{\tilde\tau_1}=327$ GeV, $m_{\tilde\tau_2}=377$ GeV for $\tgb=30$.}.
Due to the Coulomb singularities the stop and sbottom thresholds develop
spikes corresponding to the logarithmic singularities according to
Eq.~(\ref{eq:coul}).

The relative QCD corrections to the photonic Higgs decay widths are
presented in Fig.~\ref{fg:dhgaga} for the two cases, in which SUSY
particles have been taken into account or not. The QCD corrections reach
a size of 10--20\% for moderate and large Higgs masses apart from the
threshold regions, where the perturbative results are unrealiable due to
the Coulomb singularities. Since at a $\gamma\gamma$ collider the photon
fusion cross section can be measured with an accuracy of a few per cent,
these corrections have to be taken into account properly. The size of
the QCD corrections with and without SUSY particle loops is of the same
order of magnitude, but they can be of opposite sign.

In order to quantify the size of squark mass effects beyond the heavy
squark mass limit in the relative QCD corrections, the ratio between the
fully massive photonic decay width $\Gamma(h/H\to\gamma\gamma)$ at NLO
of Eqs.~(\ref{eq:hgagalo}, \ref{eq:gagaamp}) and the approximate width
$\Gamma_\infty$, where the heavy squark limit of Eq.~(\ref{eq:csqlim})
has been inserted for the coefficient $C_{\tilde Q}^{h/H}$ instead of
the fully massive result, is shown in Fig.~\ref{fg:mhgaga}. (The full
squark mass dependence of the LO form factors has been taken into
account in both expressions.) The squark mass effects reach a size of up
to $\sim 30\%$, thus underlining the relevance of including the full
mass dependences.

\subsection{Gluonic Scalar Higgs Decays}
The gluonic Higgs boson decays $h/H\to gg$ are mediated by quark and
squark triangle loops (see Fig.~\ref{fg:lodiahgg}). The lowest order
decay widths of the scalar MSSM Higgs boson decays are given by
\cite{cxn,gghnlo}
\begin{eqnarray}
\Gamma_{LO}(h/H\to gg) & = &
\frac{G_F\alpha_s^2(\mu_R) M_{h/H}^3}{36\sqrt{2}\pi^3}
\left|\sum_Q g_Q^{h/H} A_Q^{h/H}(\tau_Q) + \sum_{\tilde Q} g_{\tilde
Q}^{h/H} A_{\tilde Q}^{h/H} (\tau_{\tilde Q}) \right|^2
\label{eq:hglgllo} \\
A_Q^{h/H}(\tau) & = & \frac{3}{2} \tau [1+(1-\tau)f(\tau)] \nonumber \\
A_{\tilde Q}^{h/H} (\tau) & = & -\frac{3}{4} \tau[1-\tau
f(\tau)] \nonumber
\end{eqnarray}
using the same notation as for the photonic Higgs boson decays. For
large loop particle masses the form factors approach constant values,
\begin{eqnarray*}
A_Q^{h/H} (\tau) & \to & 1 \hspace*{1cm}
\mbox{for $M_{h/H}^2 \ll 4m_Q^2$} \nonumber \\
A_{\widetilde{Q}}^{h/H} (\tau) & \to & \frac{1}{4} \hspace*{1cm}
\mbox{for $M_{h/H}^2 \ll 4m_{\widetilde{Q}}^2$}
\, .
\end{eqnarray*}
The squark contributions become significant for squark masses below
about 400 GeV.

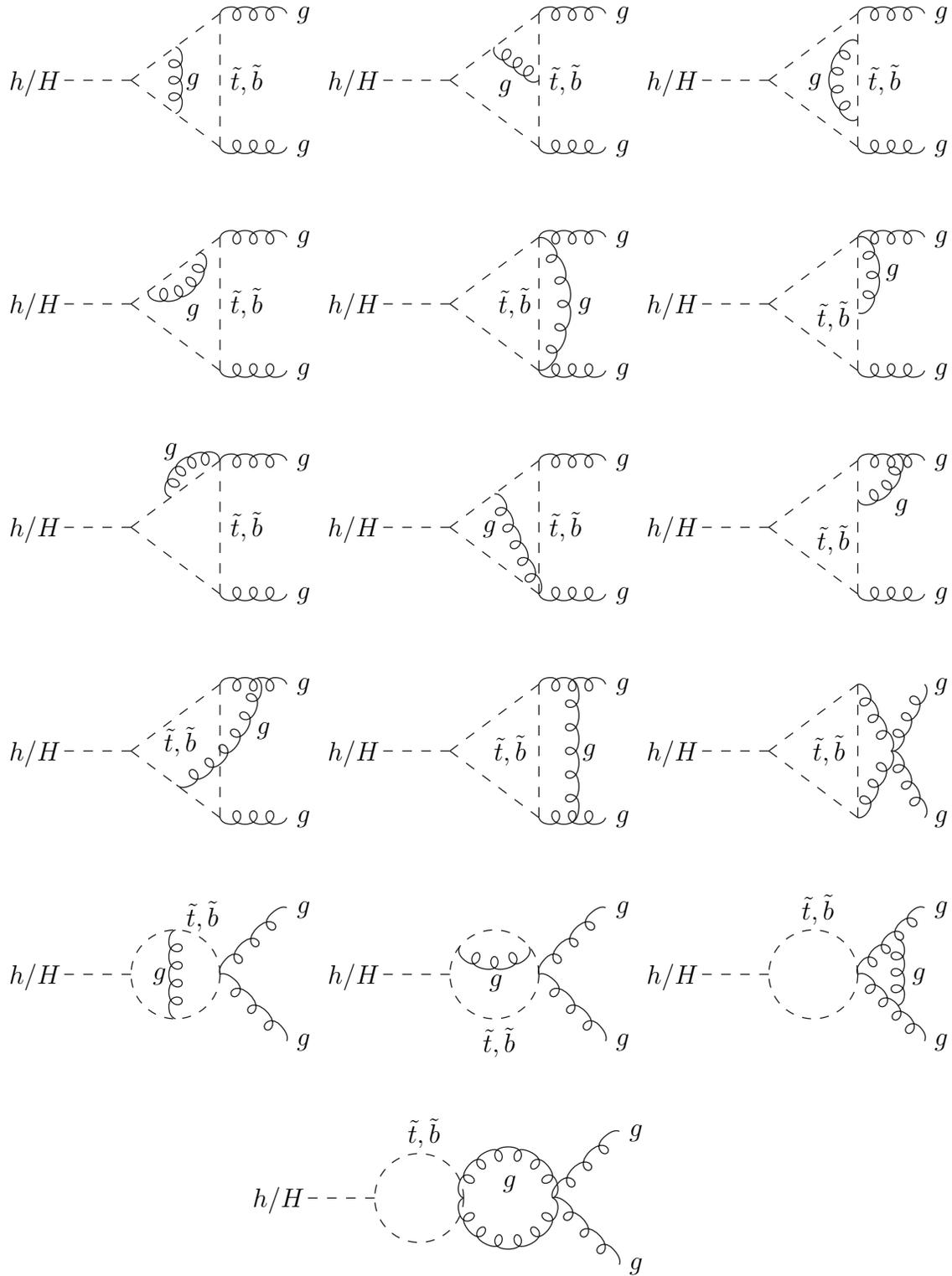
\begin{figure}[hbtp]
\begin{picture}(100,90)(-40,0)
\Gluon(70,20)(100,20){-3}{3}
\Gluon(70,80)(100,80){3}{3}
\Gluon(50,35)(50,65){-3}{3}
\DashLine(70,20)(70,80){5}
\DashLine(70,80)(30,50){5}
\DashLine(30,50)(70,20){5}
\DashLine(0,50)(30,50){5}
\put(-25,46){$h/H$}
\put(75,46){$\tilde t,\tilde b$}
\put(105,18){$g$}
\put(105,78){$g$}
\put(55,48){$g$}
\end{picture}
\begin{picture}(100,90)(-80,0)
\Gluon(70,20)(100,20){-3}{3}
\Gluon(70,80)(100,80){3}{3}
\Gluon(70,50)(50,65){3}{3}
\DashLine(70,20)(70,80){5}
\DashLine(70,80)(30,50){5}
\DashLine(30,50)(70,20){5}
\DashLine(0,50)(30,50){5}
\put(-25,46){$h/H$}
\put(75,46){$\tilde t,\tilde b$}
\put(105,18){$g$}
\put(105,78){$g$}
\put(52,45){$g$}
\end{picture}
\begin{picture}(100,90)(-120,0)
\Gluon(70,20)(100,20){-3}{3}
\Gluon(70,80)(100,80){3}{3}
\GlueArc(80,50)(20,120,240){3}{4}
\DashLine(70,20)(70,80){5}
\DashLine(70,80)(30,50){5}
\DashLine(30,50)(70,20){5}
\DashLine(0,50)(30,50){5}
\put(-25,46){$h/H$}
\put(75,46){$\tilde t,\tilde b$}
\put(105,18){$g$}
\put(105,78){$g$}
\put(48,48){$g$}
\end{picture} \\
\begin{picture}(100,100)(-40,0)
\Gluon(70,20)(100,20){-3}{3}
\Gluon(70,80)(100,80){3}{3}
\GlueArc(45,70)(16,243,371){3}{4}
\DashLine(70,20)(70,80){5}
\DashLine(70,80)(30,50){5}
\DashLine(30,50)(70,20){5}
\DashLine(0,50)(30,50){5}
\put(-25,46){$h/H$}
\put(75,46){$\tilde t,\tilde b$}
\put(105,18){$g$}
\put(105,78){$g$}
\put(55,45){$g$}
\end{picture}
\begin{picture}(100,100)(-80,0)
\Gluon(70,20)(100,20){-3}{3}
\Gluon(70,80)(100,80){3}{3}
\GlueArc(38,50)(43.5,-43,43){3}{5}
\DashLine(70,20)(70,80){5}
\DashLine(70,80)(30,50){5}
\DashLine(30,50)(70,20){5}
\DashLine(0,50)(30,50){5}
\put(-25,46){$h/H$}
\put(52,46){$\tilde t,\tilde b$}
\put(105,18){$g$}
\put(105,78){$g$}
\put(88,48){$g$}
\end{picture}
\begin{picture}(100,100)(-120,0)
\Gluon(70,20)(100,20){-3}{3}
\Gluon(70,80)(100,80){3}{3}
\GlueArc(52,63)(25,-43,43){3}{3}
\DashLine(70,20)(70,80){5}
\DashLine(70,80)(30,50){5}
\DashLine(30,50)(70,20){5}
\DashLine(0,50)(30,50){5}
\put(-25,46){$h/H$}
\put(52,40){$\tilde t,\tilde b$}
\put(105,18){$g$}
\put(105,78){$g$}
\put(83,63){$g$}
\end{picture} \\
\begin{picture}(100,100)(-40,0)
\Gluon(70,20)(100,20){-3}{3}
\Gluon(70,80)(100,80){3}{3}
\GlueArc(64,65)(16,69,185){3}{4}
\DashLine(70,20)(70,80){5}
\DashLine(70,80)(30,50){5}
\DashLine(30,50)(70,20){5}
\DashLine(0,50)(30,50){5}
\put(-25,46){$h/H$}
\put(75,46){$\tilde t,\tilde b$}
\put(105,18){$g$}
\put(105,78){$g$}
\put(45,83){$g$}
\end{picture}
\begin{picture}(100,100)(-80,0)
\Gluon(70,20)(100,20){-3}{3}
\Gluon(70,80)(100,80){3}{3}
\Gluon(50,65)(70,20){3}{5}
\DashLine(70,20)(70,80){5}
\DashLine(70,80)(30,50){5}
\DashLine(30,50)(70,20){5}
\DashLine(0,50)(30,50){5}
\put(-25,46){$h/H$}
\put(75,46){$\tilde t,\tilde b$}
\put(105,18){$g$}
\put(105,78){$g$}
\put(45,48){$g$}
\end{picture}
\begin{picture}(100,100)(-120,0)
\Gluon(70,20)(100,20){-3}{3}
\Gluon(70,80)(100,80){3}{3}
\GlueArc(70,80)(17.5,-90,9.5){3}{3}
\DashLine(70,20)(70,80){5}
\DashLine(70,80)(30,50){5}
\DashLine(30,50)(70,20){5}
\DashLine(0,50)(30,50){5}
\put(-25,46){$h/H$}
\put(52,40){$\tilde t,\tilde b$}
\put(105,18){$g$}
\put(105,78){$g$}
\put(87,58){$g$}
\end{picture} \\
\begin{picture}(100,100)(-40,0)
\Gluon(70,20)(100,20){-3}{3}
\Gluon(70,80)(100,80){3}{3}
\GlueArc(27,90)(60,-67,-7){3}{6}
\DashLine(70,20)(70,80){5}
\DashLine(70,80)(30,50){5}
\DashLine(30,50)(70,20){5}
\DashLine(0,50)(30,50){5}
\put(-25,46){$h/H$}
\put(45,50){$\tilde t,\tilde b$}
\put(105,18){$g$}
\put(105,78){$g$}
\put(87,58){$g$}
\end{picture}
\begin{picture}(100,100)(-80,0)
\Gluon(70,20)(100,20){-3}{3}
\Gluon(70,80)(100,80){3}{3}
\Gluon(85,82)(85,18){3}{5}
\DashLine(70,20)(70,80){5}
\DashLine(70,80)(30,50){5}
\DashLine(30,50)(70,20){5}
\DashLine(0,50)(30,50){5}
\put(-25,46){$h/H$}
\put(50,46){$\tilde t,\tilde b$}
\put(105,18){$g$}
\put(105,78){$g$}
\put(90,48){$g$}
\end{picture}
\begin{picture}(100,100)(-120,0)
\Gluon(70,20)(85,50){-3}{3}
\Gluon(85,50)(100,80){-3}{3}
\Gluon(70,80)(85,50){3}{3}
\Gluon(85,50)(100,20){3}{3}
\DashLine(70,20)(70,80){5}
\DashLine(70,80)(30,50){5}
\DashLine(30,50)(70,20){5}
\DashLine(0,50)(30,50){5}
\put(-25,46){$h/H$}
\put(50,46){$\tilde t,\tilde b$}
\put(105,18){$g$}
\put(105,78){$g$}
\end{picture} \\
\begin{picture}(100,100)(-40,0)
\Gluon(70,50)(100,20){3}{3}
\Gluon(70,50)(100,80){3}{3}
\Gluon(50,70)(50,30){-3}{4}
\DashCArc(50,50)(20,0,360){4}
\DashLine(0,50)(30,50){5}
\put(-25,46){$h/H$}
\put(55,72){$\tilde t,\tilde b$}
\put(105,18){$g$}
\put(105,78){$g$}
\put(40,48){$g$}
\end{picture}
\begin{picture}(100,100)(-80,0)
\Gluon(70,50)(100,20){3}{3}
\Gluon(70,50)(100,80){3}{3}
\GlueArc(50,75)(20,219,321){3}{3}
\DashCArc(50,50)(20,0,360){4}
\DashLine(0,50)(30,50){5}
\put(-25,46){$h/H$}
\put(45,15){$\tilde t,\tilde b$}
\put(105,18){$g$}
\put(105,78){$g$}
\put(48,45){$g$}
\end{picture}
\begin{picture}(100,100)(-120,0)
\Gluon(70,50)(100,20){3}{4}
\Gluon(70,50)(100,80){3}{4}
\Gluon(88,65.8)(88,36){3}{3}
\DashCArc(50,50)(20,0,360){4}
\DashLine(0,50)(30,50){5}
\put(-25,46){$h/H$}
\put(45,75){$\tilde t,\tilde b$}
\put(105,18){$g$}
\put(105,78){$g$}
\put(95,48){$g$}
\end{picture} \\
\begin{picture}(100,100)(-150,0)
\Gluon(110,50)(140,20){3}{3}
\Gluon(110,50)(140,80){3}{3}
\GlueArc(90,50)(20,0,180){3}{6}
\GlueArc(90,50)(20,180,360){3}{6}
\DashCArc(50,50)(20,0,360){4}
\DashLine(0,50)(30,50){5}
\put(-25,46){$h/H$}
\put(45,75){$\tilde t,\tilde b$}
\put(145,18){$g$}
\put(145,78){$g$}
\put(88,55){$g$}
\end{picture} \\[-1cm]
\caption[]{\label{fg:hggvdia} \it Generic diagrams for the virtual NLO
QCD corrections to the squark contributions to the gluonic Higgs
couplings.}
\end{figure}
\begin{figure}[hbtp]
\begin{picture}(100,115)(-30,0)
\Gluon(70,20)(100,20){-3}{3}
\Gluon(70,50)(100,50){3}{3}
\Gluon(70,80)(100,80){3}{3}
\DashLine(70,20)(70,80){5}
\DashLine(70,80)(30,50){5}
\DashLine(30,50)(70,20){5}
\DashLine(0,50)(30,50){5}
\put(-25,46){$h/H$}
\put(45,46){$\tilde t,\tilde b$}
\put(105,18){$g$}
\put(105,48){$g$}
\put(105,78){$g$}
\end{picture}
\begin{picture}(100,115)(-80,0)
\Gluon(70,20)(100,20){-3}{3}
\Gluon(70,80)(85,80){3}{1}
\Gluon(85,80)(100,100){3}{2}
\Gluon(85,80)(100,60){3}{2}
\DashLine(70,20)(70,80){5}
\DashLine(70,80)(30,50){5}
\DashLine(30,50)(70,20){5}
\DashLine(0,50)(30,50){5}
\put(-25,46){$h/H$}
\put(75,46){$\tilde t,\tilde b$}
\put(105,18){$g$}
\put(105,58){$g$}
\put(105,98){$g$}
\end{picture}
\begin{picture}(100,115)(-130,0)
\Gluon(70,50)(100,20){3}{3}
\Gluon(70,50)(100,80){3}{3}
\Gluon(50,70)(80,100){3}{3}
\DashCArc(50,50)(20,0,360){4}
\DashLine(0,50)(30,50){5}
\put(-25,46){$h/H$}
\put(45,15){$\tilde t,\tilde b$}
\put(105,18){$g$}
\put(105,78){$g$}
\put(85,102){$g$}
\end{picture} \\
\begin{picture}(100,100)(-30,0)
\Gluon(70,50)(100,20){3}{4}
\Gluon(70,50)(85,65){3}{2}
\Gluon(85,65)(100,80){3}{2}
\Gluon(85,65)(100,50){3}{2}
\DashCArc(50,50)(20,0,360){4}
\DashLine(0,50)(30,50){5}
\put(-25,46){$h/H$}
\put(45,15){$\tilde t,\tilde b$}
\put(105,18){$g$}
\put(105,48){$g$}
\put(105,78){$g$}
\end{picture}
\begin{picture}(100,90)(-80,0)
\Gluon(70,20)(100,20){-3}{3}
\Gluon(70,80)(85,80){3}{1}
\ArrowLine(85,80)(100,100)
\ArrowLine(100,60)(85,80)
\DashLine(70,20)(70,80){5}
\DashLine(70,80)(30,50){5}
\DashLine(30,50)(70,20){5}
\DashLine(0,50)(30,50){5}
\put(-25,46){$h/H$}
\put(75,46){$\tilde t,\tilde b$}
\put(105,18){$g$}
\put(105,58){$\bar q$}
\put(105,98){$q$}
\end{picture}
\begin{picture}(100,100)(-130,0)
\Gluon(70,50)(100,20){3}{3}
\Gluon(70,50)(85,65){3}{2}
\ArrowLine(85,65)(100,80)
\ArrowLine(100,50)(85,65)
\DashCArc(50,50)(20,0,360){4}
\DashLine(0,50)(30,50){5}
\put(-25,46){$h/H$}
\put(45,15){$\tilde t,\tilde b$}
\put(105,18){$g$}
\put(105,48){$\bar q$}
\put(105,78){$q$}
\end{picture} \\[-1cm]
\caption[]{\label{fg:hggrdia} \it Typical diagrams for the real NLO
QCD corrections to the squark contributions to the gluonic Higgs
decays.}
\end{figure}
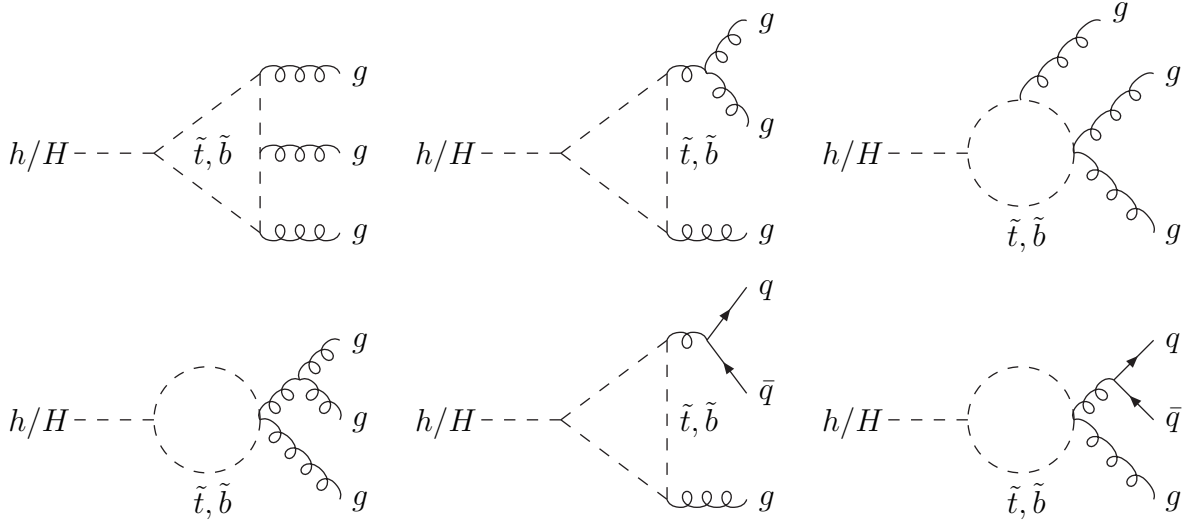
The NLO QCD corrections to the gluonic decay widths are known to be
large \cite{gghnlo,hggnlolim,hggnnlo,hggnnnlo}. They can be decomposed
into the two-loop virtual corrections and the one-loop real corrections
determined by the real radiation processes
\begin{displaymath}
h/H \to ggg \quad \mbox{and} \quad gq\bar q \, .
\end{displaymath}
The generic Feynman diagrams for the squark loop contributions are
depicted in Figs.~\ref{fg:hggvdia} and \ref{fg:hggrdia}. As in the
photonic case the Feynman integrals of the virtual corrections have been
reduced to one-dimensional integrals, which have been evaluated
numerically. In a second calculation the results have been obtained
purely numerically. Both results are in mutual agreement.

The strong coupling constant $\alpha_s$ is renormalized in the
$\overline{\rm MS}$ scheme, with the top quark and squark contributions
decoupled from the scale dependence, and the quark and squark masses are
renormalized on-shell.  The result can be cast into the form
\begin{eqnarray}
\Gamma(h/H\to gg) & = & \Gamma_{LO}(h/H\to gg) \left\{ 1 +
E^{h/H}\frac{\alpha_s}{\pi}\right\} \label{eq:Egg} \\
E^{h/H} & = & \frac{95}{4} -\frac{7}{6} N_F
+ \frac{7}{2} \Re e \left\{ \frac{
\sum_{\tilde Q} g_{\tilde Q}^{h/H} A_{\tilde Q}^{h/H} (\tau_{\tilde
Q})}{
\sum_Q g_Q^{h/H} A_Q^{h/H}(\tau_Q) +
\sum_{\tilde Q}g_{\tilde Q}^{h/H}A_{\tilde Q}^{h/H}(\tau_{\tilde
Q})}\right\} \nonumber \\
& + & \frac{33-2N_F}{6} \log \frac{\mu_R^2}{M_{h/H}^2} + \Delta E^{h/H}
\, , \nonumber
\end{eqnarray}
where the correction $\Delta E^{h/H}$ denotes the Higgs, quark and
squark mass dependent part, while the rest represents the total result
in the limit of heavy loop particle masses. The latter can be derived
from low-energy theorems analogously to the photonic case. The squark
contributions to the gluonic Higgs couplings in the limit of large
squark masses arise from the effective Lagrangian [${\cal H}=h,H$]
\begin{equation}
\Delta {\cal L}_{eff} = g_{\tilde Q}^{\cal H} \frac{1}{4} \frac{\beta_{\tilde
Q}(\alpha_s)/\alpha_s}{1+ \gamma_{m_{\tilde Q}}(\alpha_s)} G^{a\mu\nu}
G^a_{\mu\nu} \frac{\cal H}{v} \, ,
\end{equation}
where $\beta_{\tilde Q}(\alpha_s)/\alpha_s = (\alpha_s/12\pi)
[1+11\alpha_s/2\pi+\cdots]$ denotes the heavy squark contribution to the
QCD $\beta$ function and $\gamma_{m_{\tilde Q}} (\alpha_s)= \alpha_s/\pi
+ \cdots$ the anomalous squark mass dimension. The NLO expansion of the
effective Lagrangian reads as\footnote{The value for the QCD corrections
in the heavy squark limit differs from the result obtained in
Ref.~\cite{gghnlosq}. The difference can be traced back to a wrong
expression for the anomalous squark mass dimension used in
\cite{gghnlosq}.}
\begin{equation}
\Delta {\cal L}_{eff} = g_{\tilde Q}^{\cal H} \frac{\alpha_s}{48\pi}
G^{a\mu\nu} G^a_{\mu\nu} \frac{\cal H}{v}
\left[1 + \frac{9}{2} \frac{\alpha_s}{\pi} + {\cal O}(\alpha_s^2) \right] \, .
\label{eq:leffhglglsq}
\end{equation}
Including the quark contributions in the heavy quark limit, the total
effective Lagrangian is given by
\begin{equation}
{\cal L}_{eff} = \frac{\alpha_s}{12\pi}
G^{a\mu\nu} G^a_{\mu\nu} \frac{\cal H}{v}
\left\{\sum_Q g_Q^{\cal H} \left[1+\frac{11}{4}\frac{\alpha_s}{\pi}\right] +
\sum_{\tilde Q} \frac{g_{\tilde Q}^{\cal H}}{4} \left[1 + \frac{9}{2}
\frac{\alpha_s}{\pi}\right] + {\cal O}(\alpha_s^2) \right\} \, ,
\label{eq:leffhglgl}
\end{equation}
The calculation based on this effective Lagrangian yields the following
results for the finite parts of the individual contributions to the
coefficients $E^{h/H}$,
\begin{eqnarray}
E^{h/H} & = & E_{virt}^{h/H} + E_{ggg}^{h/H} + E_{gq\bar q}^{h/H} +
\frac{33-2N_F}{6} \log \frac{\mu_R^2}{M_{h/H}^2} \non \\
E_{virt}^{h/H} & = & \pi^2 + \frac{11}{2} + \frac{7}{2} \Re e
\left\{ \frac{ \sum_{\tilde Q} g_{\tilde Q}^{h/H} A_{\tilde Q}^{h/H}
(\tau_{\tilde Q})}{ \sum_Q g_Q^{h/H} A_Q^{h/H}(\tau_Q) + \sum_{\tilde
Q}g_{\tilde Q}^{h/H}A_{\tilde Q}^{h/H}(\tau_{\tilde Q})}\right\}
\nonumber \\
E_{ggg}^{h/H} & = & -\pi^2 + \frac{73}{4} \non \\
E_{gq\bar q}^{h/H} & = & -\frac{7}{6} N_F \, ,
\end{eqnarray}
which agrees with the explicit results for the coefficients $E^{h/H}$ of
Eq.~(\ref{eq:Egg}) in the heavy quark and squark limits, where $\Delta
E^{h/H}$ vanishes.

Since at the $\tilde Q\bar{\tilde Q}$ thresholds $0^{++}$ states can
form, the NLO QCD corrections exhibit Coulomb singularities as for the
photonic Higgs couplings. The singular behavior can be derived from the
Sommerfeld rescattering corrections and leads to the following
expressions at each specific $\tilde Q_0 \bar{\tilde Q}_0$ threshold,
\begin{equation}
E^{h/H} \to \Re e \left\{ \frac{g^{h/H}_{\tilde Q_0} A^{h/H}_{\tilde Q_0}
(\tau_{\tilde Q_0}) \frac{16\pi^2}{3(\pi^2-4)} \left[ -\log(\tau_{\tilde
Q_0}^{-1}-1) + i\pi + const \right]}{\sum_Q g_Q^{h/H} A_Q^{h/H} (\tau_Q)
+ \sum_{\tilde Q} g_{\tilde Q}^{h/H} A_{\tilde Q}^{h/H} (\tau_{\tilde
Q})} \right\} \, ,
\end{equation}
which agrees quantitatively with the numerical results.

\begin{figure}[hbtp]
\begin{picture}(100,500)(0,0)
\put(40.0,120.0){\includegraphics{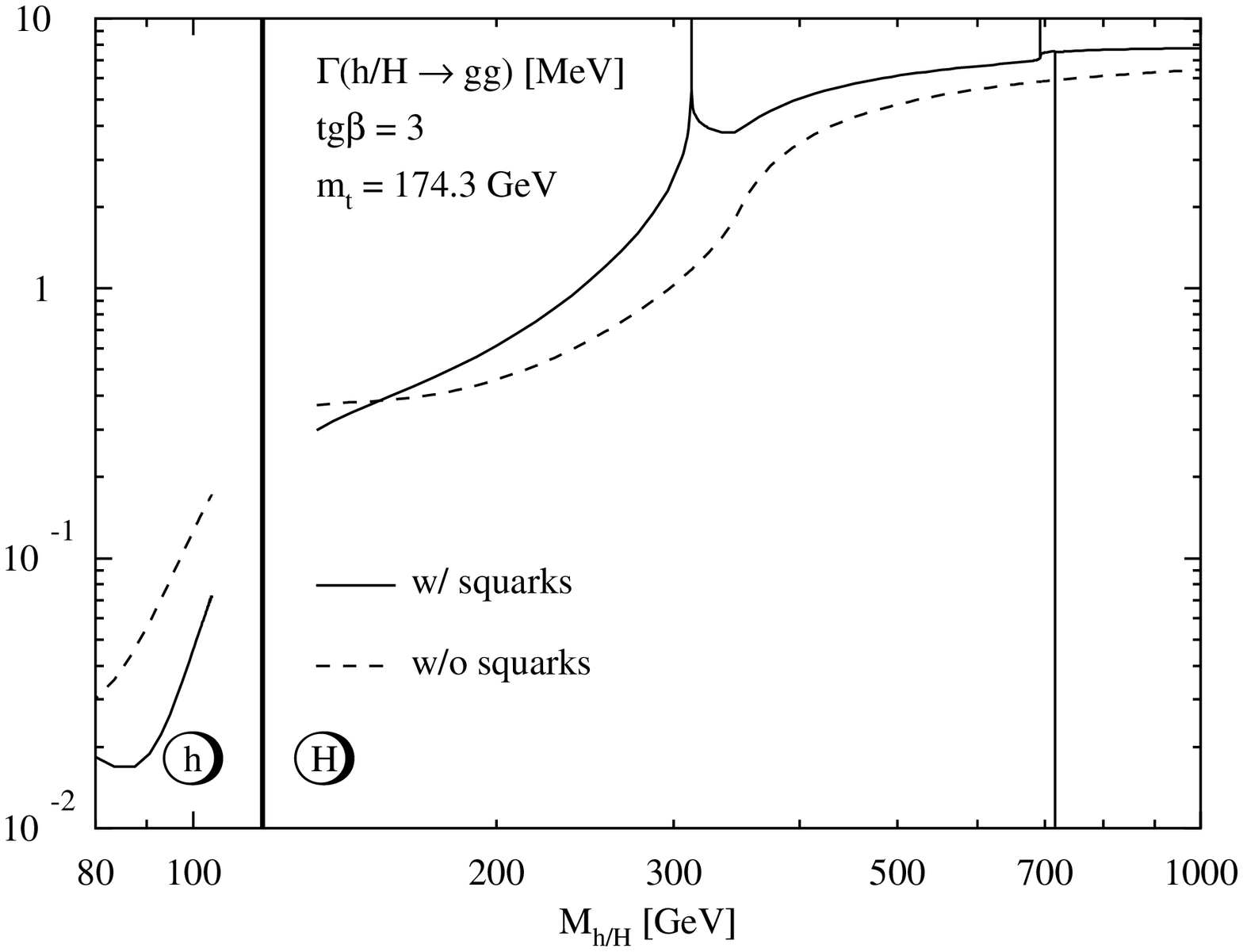}}
\put(40.0,-135.0){\includegraphics{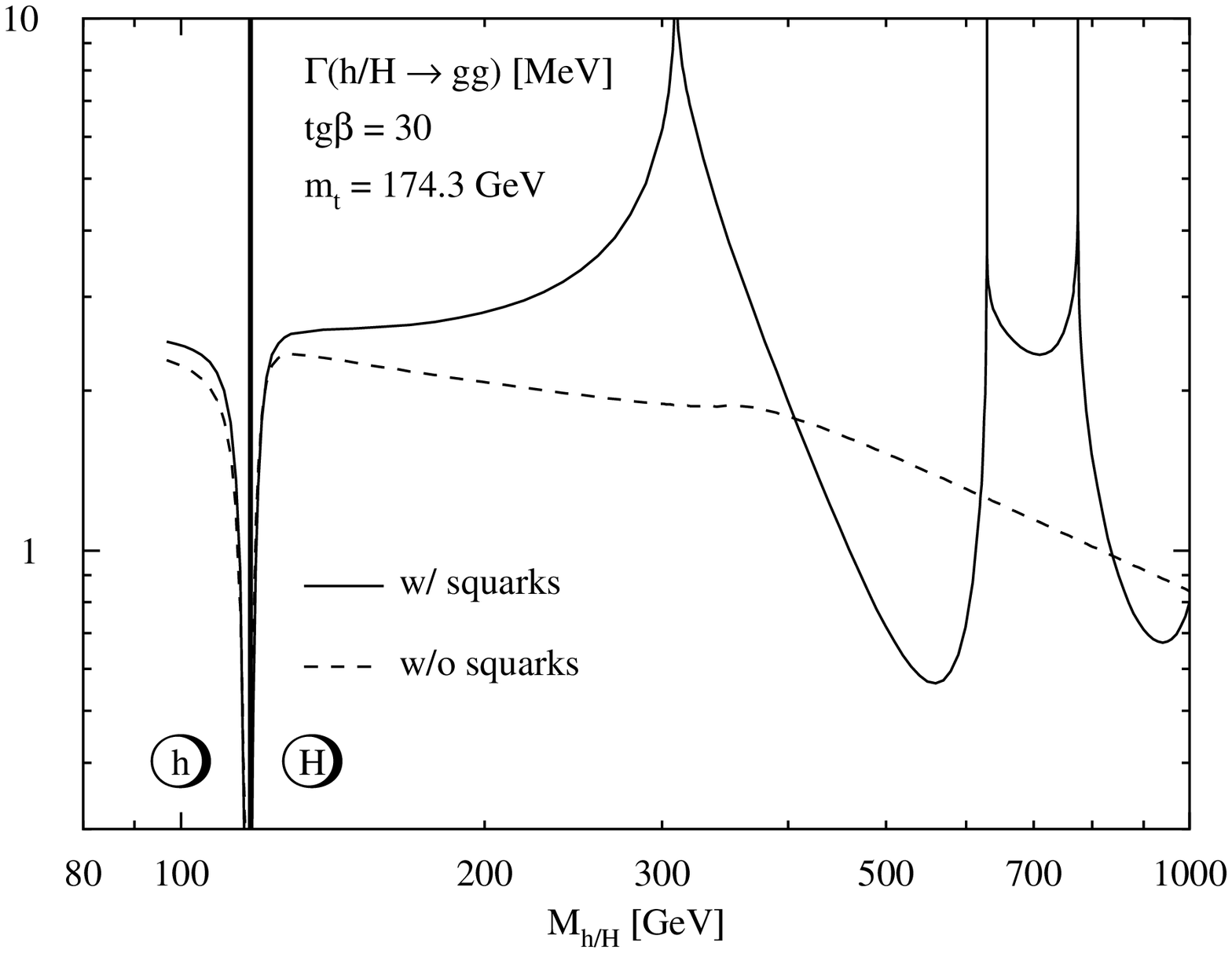}}
\end{picture}
\caption[]{\label{fg:hglglqsq} \it QCD corrected partial decay widths of
the scalar MSSM Higgs bosons to two gluons as functions of the
corresponding Higgs masses for $\tgb=3$ and 30. The full curves include
all loop contributions, while in the dashed lines the squark
contributions are omitted.  The kinks, bumps and spikes correspond to the
$\tilde t_1\bar{\tilde t}_1, t\bar t, \tilde b_1\bar{\tilde b}_1$ and $\tilde
b_2\bar{\tilde b}_2$ thresholds in consecutive order with rising Higgs
mass. The renormalization scale of the strong coupling $\alpha_s$ is
chosen as the corresponding Higgs mass.}
\end{figure}

The partial decay widths into gluons are presented in
Fig.~\ref{fg:hglglqsq} with and without the squark contributions for two
values of $\tgb=3,30$ in the gluophobic Higgs scenario. The
renormalization scale has been identified with the corresponding Higgs
mass, $\mu_R=M_{h/H}$. The partial widths range between $10^{-2}$ and 10
MeV. The comparison of the two curves underlines the large size of the
squark contributions, which are of the same order as the quark loops in
this scenario. The spikes of the full results originate from the Coulomb
singularities of the $\tilde t_1 \bar{\tilde t}_1, \tilde b_1
\bar{\tilde b}_1$ and $\tilde b_2 \bar{\tilde b}_2$ thresholds in
consecutive order.

\begin{figure}[hbtp]
\begin{picture}(100,500)(0,0)
\put(40.0,120.0){\includegraphics{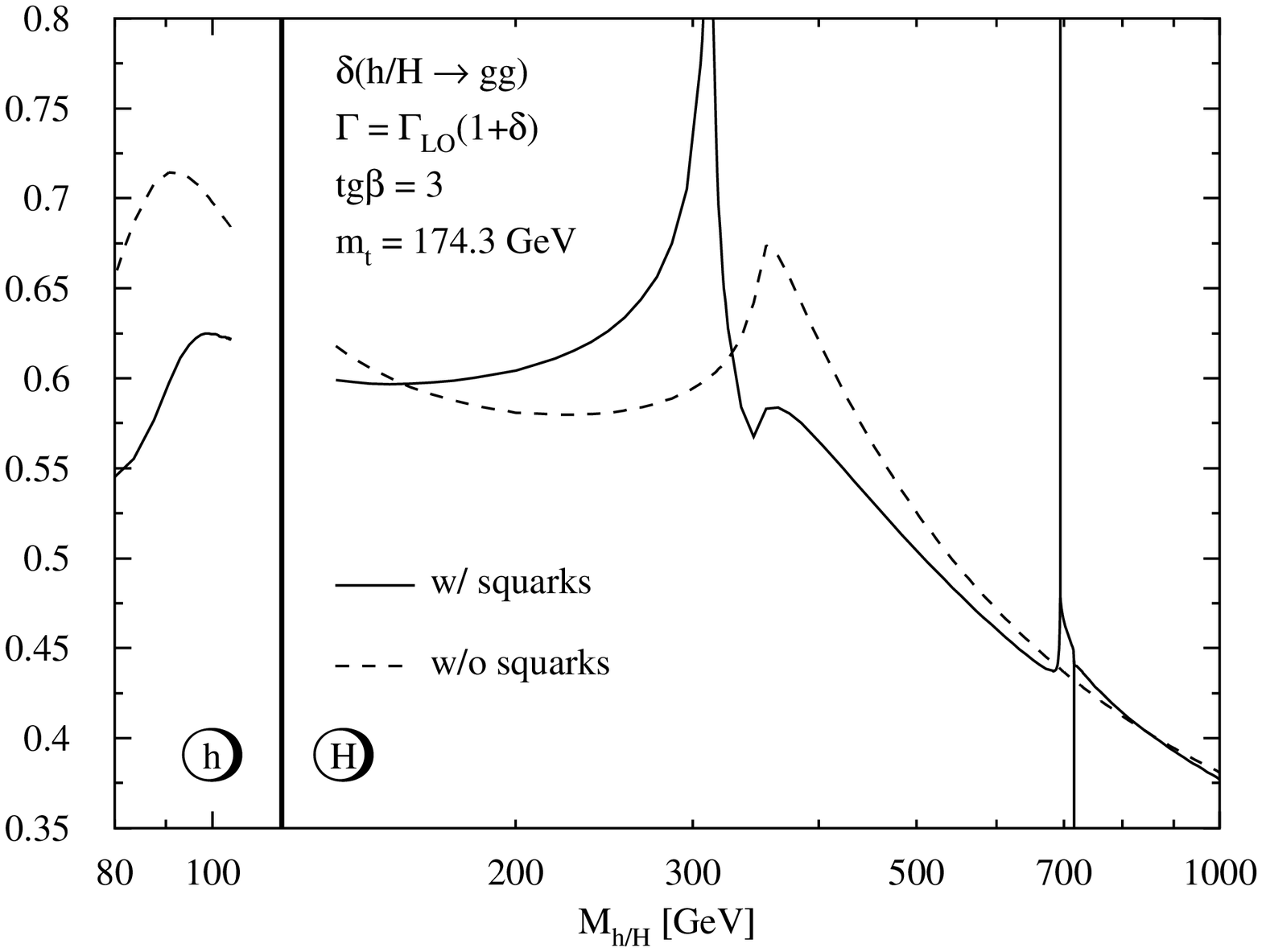}}
\put(40.0,-135.0){\includegraphics{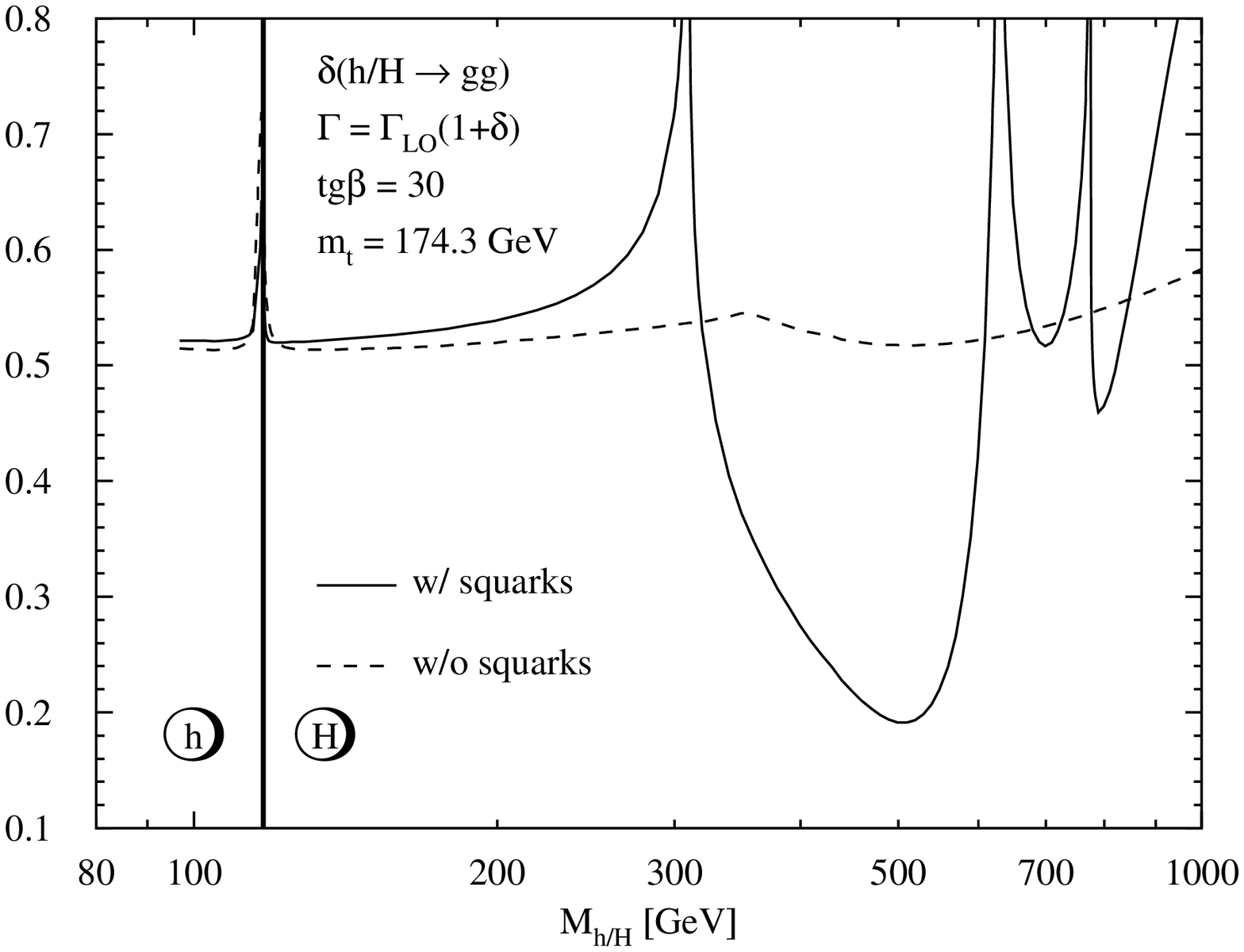}}
\end{picture}
\caption[]{\label{fg:hglgldel} \it Relative QCD corrections to the
partial decay widths of the scalar MSSM Higgs bosons to two gluons as
functions of the corresponding Higgs masses for $\tgb=3$ and 30. The
full curves include all loop contributions, while in the dashed lines
the squark contributions are omitted. The kinks, bumps and spikes correspond to
the $\tilde t_1\bar{\tilde t}_1, t\bar t, \tilde b_1\bar{\tilde b}_1$
and $\tilde b_2\bar{\tilde b}_2$ thresholds in consecutive order with
rising Higgs mass.  The renormalization scale of the strong coupling
$\alpha_s$ is chosen as the corresponding Higgs mass.}
\end{figure}

Fig.~\ref{fg:hglgldel} displays the relative QCD corrections to the
gluonic Higgs decays with and without squark contributions. Apart from
the threshold singularities they are of similar size and increase the
decay widths by ${\cal O}(50\%)$.

\begin{figure}[hbtp]
\begin{picture}(100,500)(0,0)
\put(40.0,120.0){\includegraphics{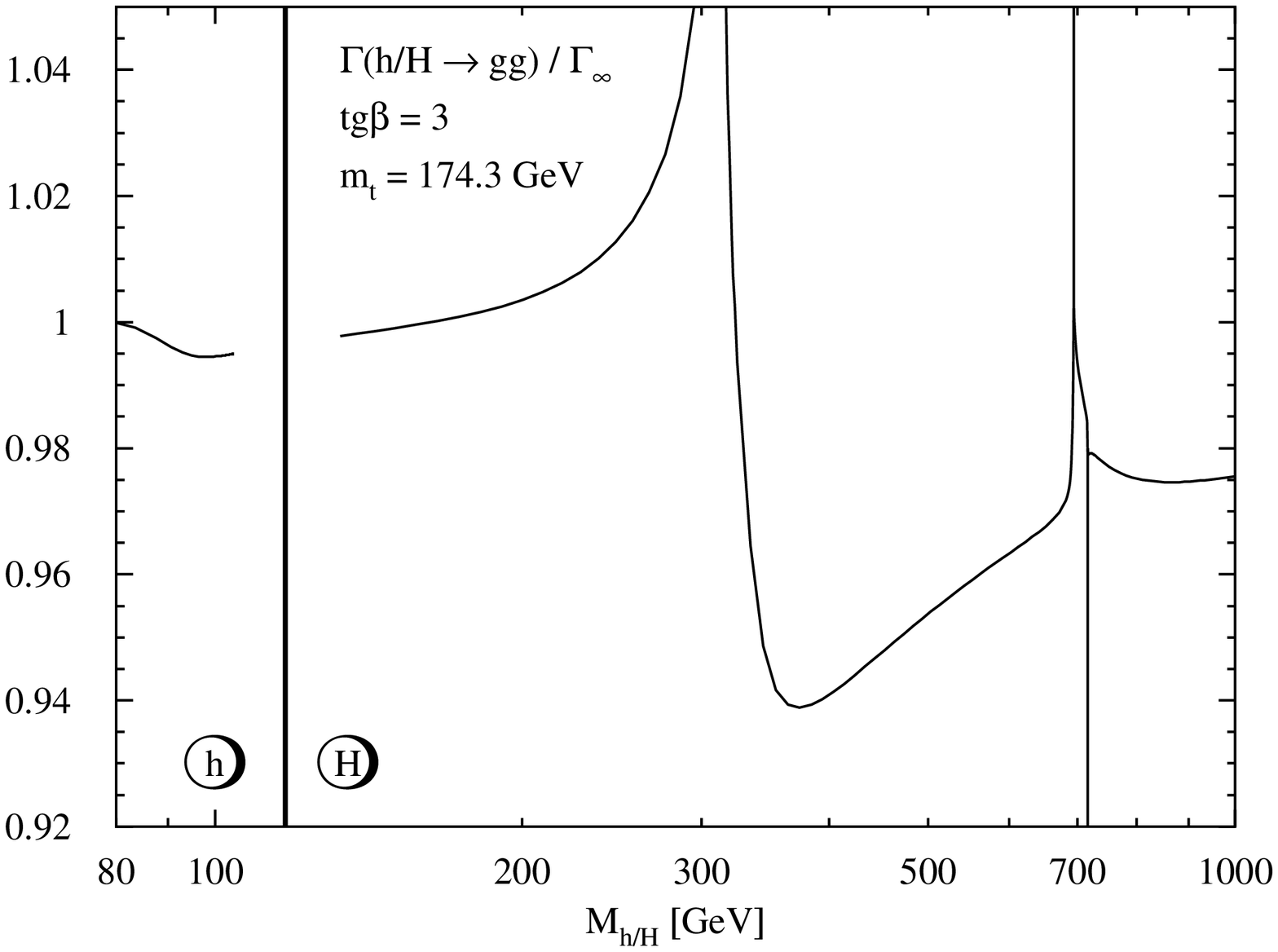}}
\put(40.0,-135.0){\includegraphics{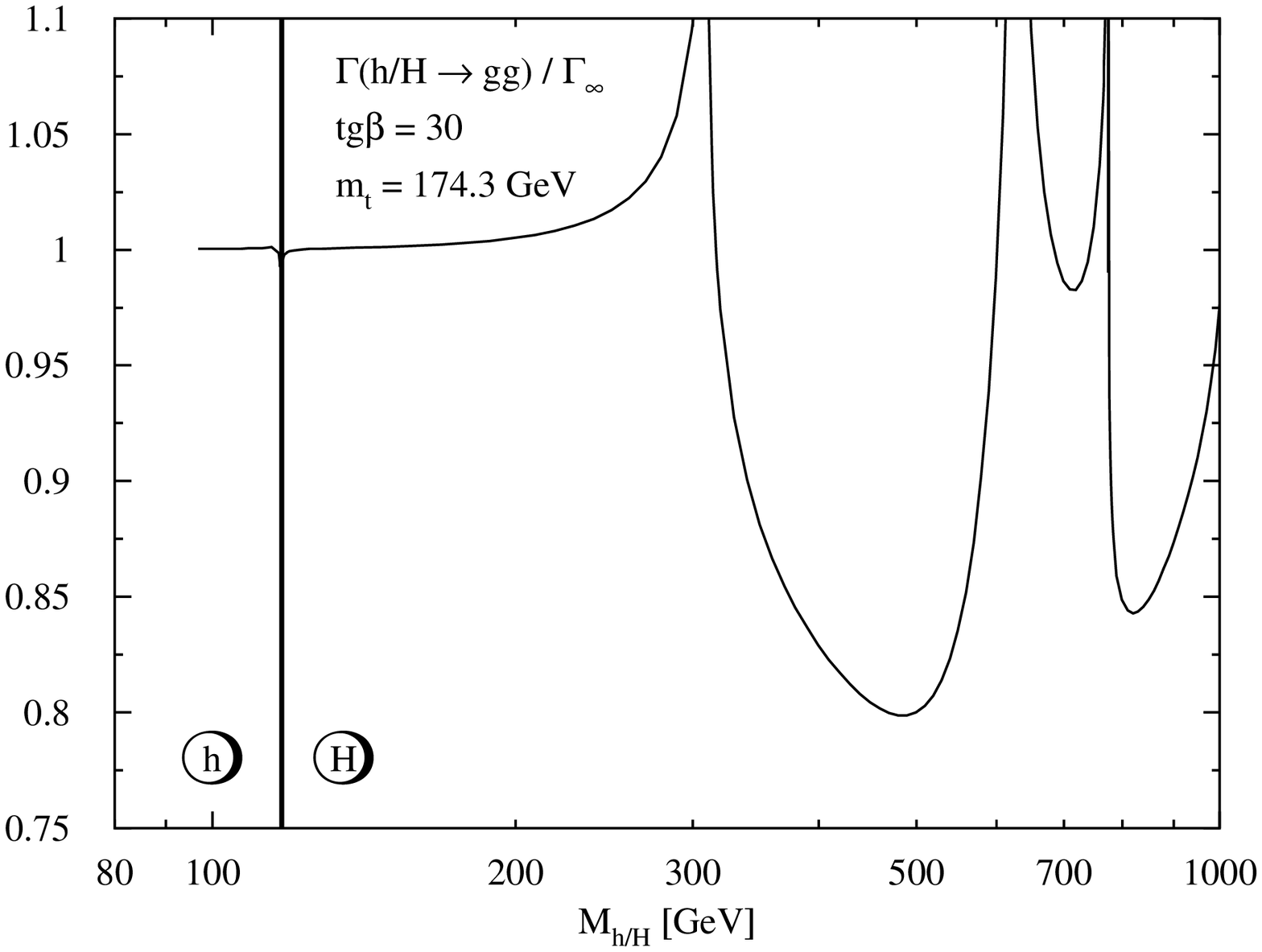}}
\end{picture}
\caption[]{\label{fg:hglglmass} \it Ratio of the QCD corrected partial
decay widths of the scalar MSSM Higgs bosons to two gluons including the
full squark mass dependence and those obtained by taking the relative
QCD corrections to the squark loops in the heavy mass limit as functions
of the corresponding Higgs masses for $\tgb=3$ and 30.  The kinks and
spikes correspond to the $\tilde t_1\bar{\tilde t}_1, \tilde
b_1\bar{\tilde b}_1$ and $\tilde b_2\bar{\tilde b}_2$ thresholds in
consecutive order with rising Higgs mass. The renormalization scale of
the strong coupling $\alpha_s$ is chosen as the corresponding Higgs
mass.}
\end{figure}

In order to quantify the squark mass effects on the relative QCD
corrections, Fig.~\ref{fg:hglglmass} depicts the ratio between the fully
massive gluonic decay widths $\Gamma(h/H\to gg)$ and the widths obtained
by taking the squark contributions to the coefficients $E^{h/H}$ in the
infinite squark mass limits, while the LO decay widths are used with the full
squark mass dependence. It is clearly visible that squark mass effects on
the relative NLO corrections can reach 20--30\% in addition to the squark
mass dependence at LO. Thus the fully massive results are significant
for a proper quantitative prediction of the gluonic decay widths at NLO.
The heavy squark mass limit turns out to be less reliable than the heavy
top mass limit for the top quark loops.

\subsection{Gluon Fusion}
The gluon fusion processes $gg\to h/H$ are mediated by heavy quark and
squark triangle loops with the latter contributing significantly for
squark masses below 400 GeV. The lowest order cross sections in the
narrow-width approximation can be obtained from the gluonic decay widths
of the scalar Higgs bosons \cite{cxn,gghnlo,gghlo},
\begin{eqnarray}
\sigma_{LO}(pp\to h/H) & = & \sigma^{h/H}_0 \tau_{h/H}\frac{d{\cal
L}^{gg}}{d\tau_{h/H}} \\
\sigma^{h/H}_0 & = & \frac{\pi^2}{8M_{h/H}^3}\Gamma_{LO}(h/H\to gg)
\nonumber \\
\sigma^{h/H}_0 & = & \frac{G_{F}\alpha_{s}^{2}(\mu_R)}{288 \sqrt{2}\pi} \
\left| \sum_{Q} g_Q^{h/H} A_Q^{h/H} (\tau_{Q})
+ \sum_{\widetilde{Q}} g_{\widetilde{Q}}^{h/H} A_{\widetilde{Q}}^{h/H}
(\tau_{\widetilde{Q}}) \right|^{2} \, , 
\end{eqnarray}
where $\tau_{h/H} = M_{h/H}^2/s$ with $s$ specifying the squared hadronic
c.m.\ energy. The LO form factors are identical to the corresponding form
factors $A^{h/H}_{Q/\tilde Q}$ of the gluonic decay modes,
Eq.~(\ref{eq:hglgllo}). The gluon luminosity at the factorization scale
$\mu_F$ is defined as
\begin{displaymath}
\frac{d{\cal L}^{gg}}{d\tau} = \int_\tau^1 \frac{dx}{x}~g(x,\mu_F^2)
g(\tau /x,\mu_F^2) \, ,
\end{displaymath}
where $g(x,\mu_F^2)$ denotes the gluon parton density of the proton.

\begin{figure}[htbp]
\begin{center}
\begin{picture}(130,100)(20,0)
\Gluon(10,20)(50,20){-3}{4}
\Gluon(10,80)(50,80){3}{4}
\DashLine(50,20)(50,80){5}
\DashLine(50,80)(90,80){5}
\DashLine(90,80)(90,20){5}
\DashLine(90,20)(50,20){5}
\Gluon(90,80)(130,80){3}{4}
\DashLine(90,20)(130,20){5}
\put(0,18){$g$}
\put(0,78){$g$}
\put(30,46){$\tilde t,\tilde b$}
\put(135,78){$g$}
\put(100,25){$h,H$}
\end{picture}
\begin{picture}(130,100)(0,0)
\Gluon(10,20)(50,20){-3}{4}
\Gluon(10,80)(50,80){3}{4}
\Gluon(30,83)(70,100){3}{4}
\DashLine(50,20)(50,80){5}
\DashLine(50,80)(90,50){5}
\DashLine(90,50)(50,20){5}
\DashLine(90,50)(130,50){5}
\put(0,18){$g$}
\put(0,78){$g$}
\put(75,98){$g$}
\put(30,46){$\tilde t,\tilde b$}
\put(100,36){$h,H$}
\end{picture}
\begin{picture}(130,100)(-20,0)
\Gluon(10,20)(50,50){-3}{5}
\Gluon(10,80)(50,50){3}{5}
\Gluon(31,66.5)(70,95){3}{5}
\DashCArc(70,50)(20,180,360){5}
\DashCArc(70,50)(20,0,180){5}
\DashLine(90,50)(130,50){5}
\put(0,18){$g$}
\put(0,78){$g$}
\put(75,93){$g$}
\put(60,15){$\tilde t,\tilde b$}
\put(100,36){$h,H$}
\end{picture}  \\
\begin{picture}(130,110)(20,0)
\Gluon(10,20)(50,20){-3}{4}
\Gluon(30,80)(50,80){3}{2}
\ArrowLine(10,80)(30,80)
\ArrowLine(30,80)(70,110)
\DashLine(50,20)(50,80){5}
\DashLine(50,80)(90,50){5}
\DashLine(90,50)(50,20){5}
\DashLine(90,50)(130,50){5}
\put(0,18){$g$}
\put(0,78){$q$}
\put(75,108){$q$}
\put(30,46){$\tilde t,\tilde b$}
\put(100,36){$h,H$}
\end{picture}
\begin{picture}(130,110)(0,0)
\ArrowLine(10,80)(50,50)
\ArrowLine(50,50)(10,20)
\Gluon(50,50)(90,50){3}{4}
\Gluon(130,80)(170,80){3}{4}
\DashLine(90,50)(130,80){5}
\DashLine(90,50)(130,20){5}
\DashLine(130,80)(130,20){5}
\DashLine(130,20)(170,20){5}
\put(0,18){$\bar q$}
\put(0,78){$q$}
\put(70,58){$g$}
\put(135,46){$\tilde t,\tilde b$}
\put(140,25){$h,H$}
\end{picture} \\
\caption[]{\label{fg:gghrdia} \it Typical diagrams for the real NLO
QCD corrections to the squark contributions to the gluon fusion
processes.}
\end{center}
\end{figure}
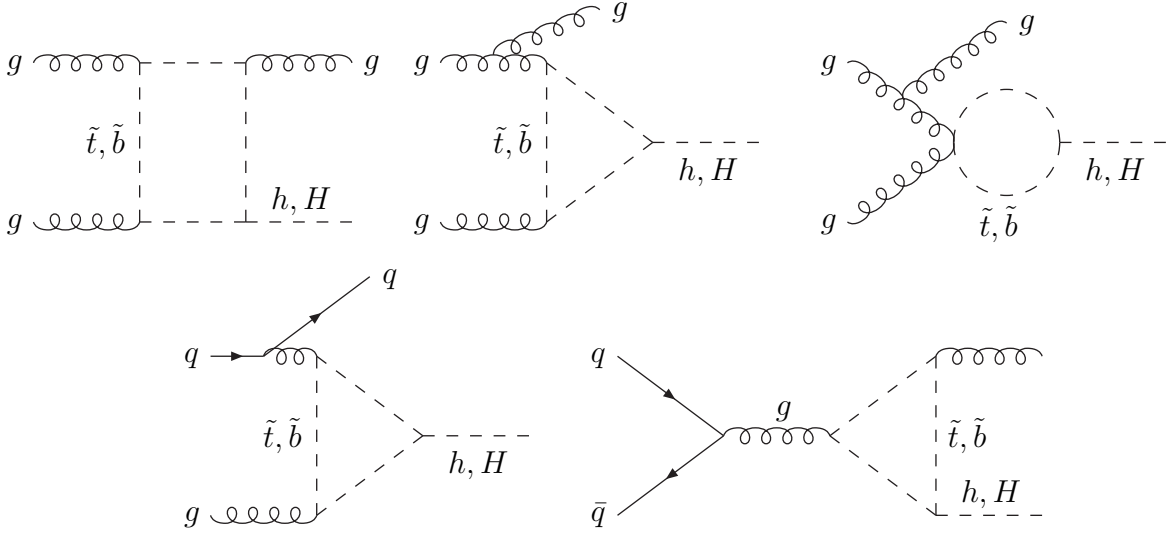
The NLO QCD corrections consist of the virtual two-loop corrections,
corresponding to the diagrams of Fig.~\ref{fg:hggvdia}, as well as the
real corrections due to the radiation processes $gg\to gh/H, gq\to qh/H$
and $q\bar q\to gh/H$. The diagrams of the real corrections are shown in
Fig.~\ref{fg:gghrdia}. While the Higgs bosons do not acquire any
transverse momentum at LO, they appear at finite transverse momenta in
these radiation processes corresponding to Higgs + jet production. The
relevance of quark mass effects on the shapes of the transverse momentum
distributions has been demonstrated in Ref.~\cite{pthiggs}, so that
similar effects may be expected for the squark mass dependence.  The
final results for the total hadronic cross sections can be split into five
parts accordingly, 
\begin{equation} 
\sigma(pp \rightarrow h/H+X) = \sigma^{h/H}_{0} \left[ 1+ C^{h/H}
\frac{\alpha_{s}}{\pi} \right] \tau_{h/H} \frac{d{\cal
L}^{gg}}{d\tau_{h/H}} + \Delta \sigma^{h/H}_{gg} + \Delta
\sigma^{h/H}_{gq} + \Delta \sigma^{h/H}_{q\bar{q}} \, ,
\label{eq:glufuscxn}
\end{equation}
The strong coupling constant is renormalized in the $\overline{\rm MS}$
scheme, with the top quark and squark contributions decoupled from the
scale dependence. The quark and squark masses are renormalized
on-shell. The parton densities are defined in the $\overline{\rm MS}$
scheme with five active flavors, i.e. the top quark and the squarks are
not included in the factorization scale dependence. The virtual
coefficients split into the infrared $\pi^2$ term, a logarithmic term
including the renormalization scale $\mu_R$ and finite (s)quark mass
dependent pieces $c^{h/H}(\tau_Q,\tau_{\tilde Q})$,
\begin{equation}
C^{h/H}(\tau_Q,\tau_{\tilde Q}) = \pi^{2}+ c^{h/H}(\tau_Q,\tau_{\tilde
Q}) + \frac{33-2N_{F}}{6} \log
\frac{\mu_R^{2}}{M_{h/H}^{2}} \, .
\label{eq:Cvirt}
\end{equation}
The finite hard contributions from gluon radiation as well as $gq$ and
$q\bar q$ scattering depend after mass factorization on the
renormalization and factorization scales $\mu_R,\mu_F$ and are given by
\begin{eqnarray}
\Delta \sigma^{h/H}_{gg} & = & \int_{\tau_{h/H}}^{1} d\tau \frac{d{\cal
L}^{gg}}{d\tau} \times \frac{\alpha_{s}}{\pi} \sigma^{h/H}_{0} \left\{ - z
P_{gg} (z) \log \frac{\mu_F^{2}}{\hat{s}} + d^{h/H}_{gg}
(z,\tau_Q,\tau_{\tilde Q}) \right. \non \\
& & \left. \hspace{3.7cm} + 12 \left[ \left(\frac{\log
(1-z)}{1-z} \right)_+ - z[2-z(1-z)] \log (1-z) \right] \right\} \non \\
\non \\
\Delta \sigma^{h/H}_{gq} & = & \int_{\tau_{H}}^{1} d\tau \sum_{q,
\bar{q}} \frac{d{\cal L} ^{gq}}{d\tau} \times \frac{\alpha_{s}}{\pi}
\sigma^{h/H}_{0} \left\{ -\frac{z}{2} P_{gq}(z)
\log\frac{\mu_F^{2}}{\hat{s}(1-z)^2}
+ d^{h/H}_{gq} (z,\tau_Q,\tau_{\tilde Q}) \right\} \non \\ \non \\
\Delta \sigma^{h/H}_{q\bar{q}} & = & \int_{\tau_{H}}^{1} d\tau
\sum_{q} \frac{d{\cal L}^{q\bar{q}}}{d\tau} \times
\frac{\alpha_{s}}{\pi}
\sigma^{h/H}_{0}~d^{h/H}_{q\bar q} (z,\tau_Q,\tau_{\tilde Q}) \, ,
\label{eq:gghqcd}
\end{eqnarray}
with $z = \tau_{h/H} / \tau = M_{h/H}^2/\hat s$, where $\hat s$ denotes
the squared partonic c.m.~energy; $P_{gg}$ and $P_{gq}$ are the
standard Altarelli--Parisi splitting functions \cite{apsplit}:
\begin{eqnarray} 
P_{gg}(z) & = & 6 \left\{ \left( \frac{1}{1-z} \right)_+ + \frac{1}{z}
-2 +
z (1-z) \right\} + \frac{33-2N_F}{6} \delta(1-z) \nonumber \\
P_{gq}(z) & = & \frac{4}{3} \frac{1+ (1-z)^2}{z} \, .
\label{eq:APKernel}
\end{eqnarray}
The natural scale choices turn out to be $\mu_R=\mu_F=M_{h/H}$.
The quark and squark mass dependences are contained in the in the
kernels $c^{h/H}(\tau_Q,\tau_{\tilde Q})$ and
$d^{h/H}_{ij}(z,\tau_Q,\tau_{\tilde Q})$ in addition to the LO
coefficients $\sigma^{h/H}_0$.  In the heavy loop particle mass limit
they reduce to simple expressions,
\begin{eqnarray}
c^{h/H}(\tau_Q,\tau_{\tilde Q}) & \to & \frac{11}{2} + \frac{7}{2} \Re e
\left\{ \frac{ \sum_{\tilde Q} g_{\tilde Q}^{h/H} A_{\tilde Q}^{h/H}
(\tau_{\tilde Q})}{ \sum_Q g_Q^{h/H} A_Q^{h/H}(\tau_Q) + \sum_{\tilde
Q}g_{\tilde Q}^{h/H}A_{\tilde Q}^{h/H}(\tau_{\tilde Q})}\right\}
\nonumber \\
d^{h/H}_{gg}(z,\tau_Q,\tau_{\tilde Q}) & \to & -\frac{11}{2} (1-z)^3
\non \\
d^{h/H}_{gq}(z,\tau_Q,\tau_{\tilde Q}) & \to & \frac{2}{3}z^2 - (1 -
z)^2 \non \\
d^{h/H}_{q\bar q}(z,\tau_Q,\tau_{\tilde Q}) & \to & \frac{32}{27}
(1-z)^3 \, .
\label{eq:gghqcdlim}
\end{eqnarray}
These can also be derived from the effective Lagrangian
Eq.~(\ref{eq:leffhglgl}), so that the full calculation agrees with the
derivation from the low-energy theorems in the heavy loop particle mass
limits.

Due to the formation of $0^{++}$ states at the $\tilde Q\bar{\tilde Q}$
thresholds the NLO QCD corrections develop Coulomb singularities
analogous to the photonic Higgs decay modes. The singular behavior can
be derived from the Sommerfeld rescattering corrections in the same way
as in the photonic case, leading to the following expressions at each
specific $\tilde Q_0 \bar{\tilde Q}_0$ threshold,
\begin{equation}
c^{h/H} \to \Re e \left\{ \frac{g^{h/H}_{\tilde Q_0} A^{h/H}_{\tilde Q_0}
(\tau_{\tilde Q_0}) \frac{16\pi^2}{3(\pi^2-4)} \left[ -\log(\tau_{\tilde
Q_0}^{-1}-1) + i\pi + const \right]}{\sum_Q g_Q^{h/H} A_Q^{h/H} (\tau_Q)
+ \sum_{\tilde Q} g_{\tilde Q}^{h/H} A_{\tilde Q}^{h/H} (\tau_{\tilde
Q})} \right\} \, ,
\end{equation}
which agrees quantitatively with the numerical results.

\begin{figure}[hbtp]
\begin{picture}(100,500)(0,0)
\put(40.0,120.0){\includegraphics{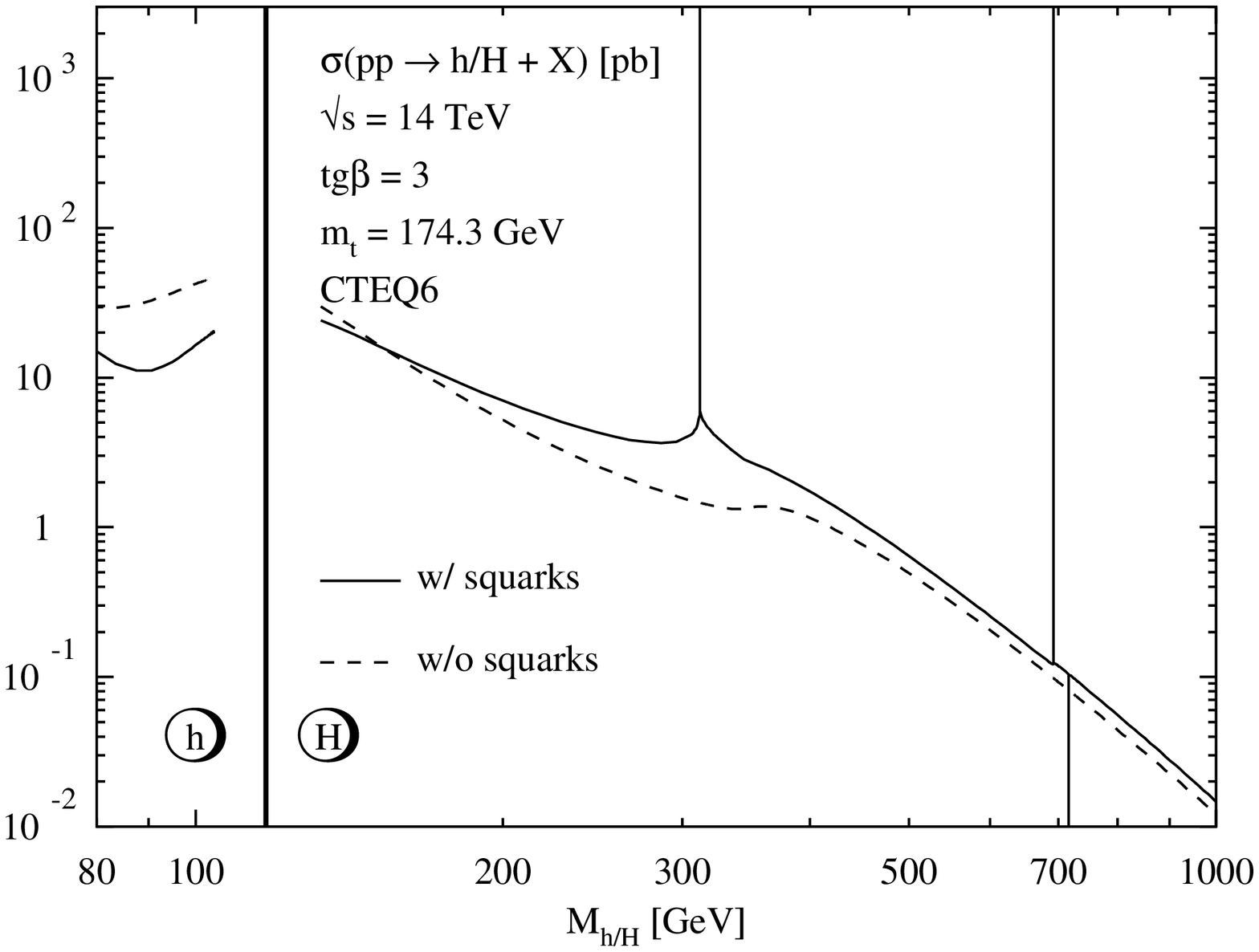}}
\put(40.0,-135.0){\includegraphics{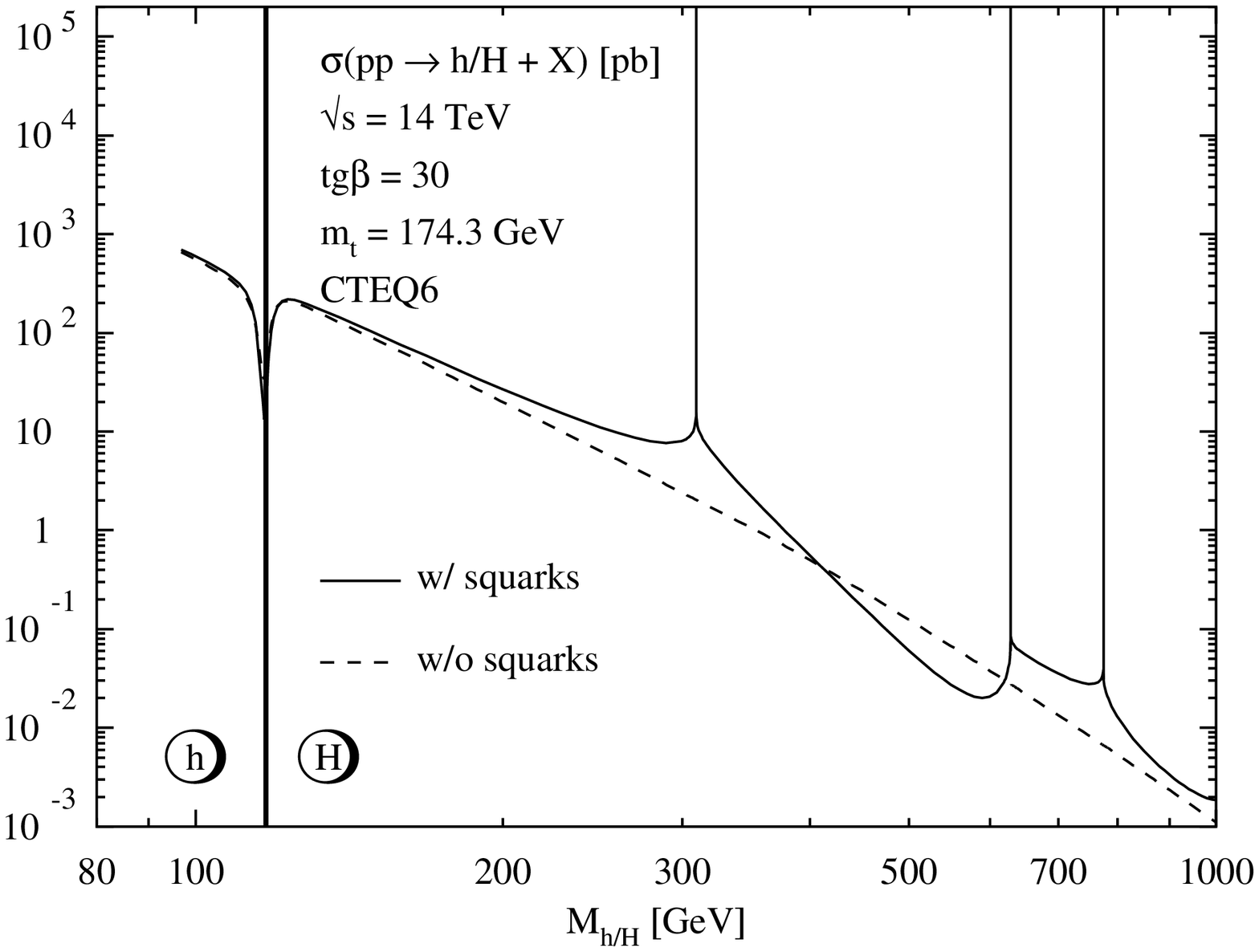}}
\end{picture}
\caption[]{\label{fg:gghqsq} \it QCD corrected production cross sections
of the scalar MSSM Higgs bosons via gluon fusion as functions of the
corresponding Higgs masses for $\tgb=3$ and 30. The full curves include
all loop contributions, while in the dashed lines the squark
contributions are omitted. The kinks, bumps and spikes correspond to the
$\tilde t_1\bar{\tilde t}_1, t\bar t, \tilde b_1\bar{\tilde b}_1$ and $\tilde
b_2\bar{\tilde b}_2$ thresholds in consecutive order with rising Higgs
mass. The renormalization and factorization scales are chosen as the
corresponding Higgs mass.}
\end{figure}

The total gluon fusion cross section at NLO is displayed in
Fig.~\ref{fg:gghqsq} with and without squark loop contributions. The
renormalization and factorization scales have been identified with the
corresponding Higgs mass, $\mu_R=\mu_F=M_{h/H}$. In the gluophobic Higgs
scenario the squark loops alter the size of the cross sections by up to
factors of about three. The spikes of the full curves originate from the
Coulomb singularities at the $\tilde t_1 \bar{\tilde t}_1, \tilde b_1
\bar{\tilde b}_1$ and $\tilde b_2 \bar{\tilde b}_2$ thresholds in
consecutive order analogously to the photonic and gluonic Higgs
couplings. It can be inferred from Fig.~\ref{fg:gghqsq} that squark
effects are always important, if the Higgs mass exceeds the
corresponding squark-antisquark threshold. This is confirmed in
particular for large values of $\tgb$, where the top quark contribution
is less important, but sizeable squark effects are visible close to and
beyond the different thresholds. This feature also holds in other MSSM
scenarios.

\begin{figure}[hbtp]
\begin{picture}(100,500)(0,0)
\put(40.0,120.0){\includegraphics{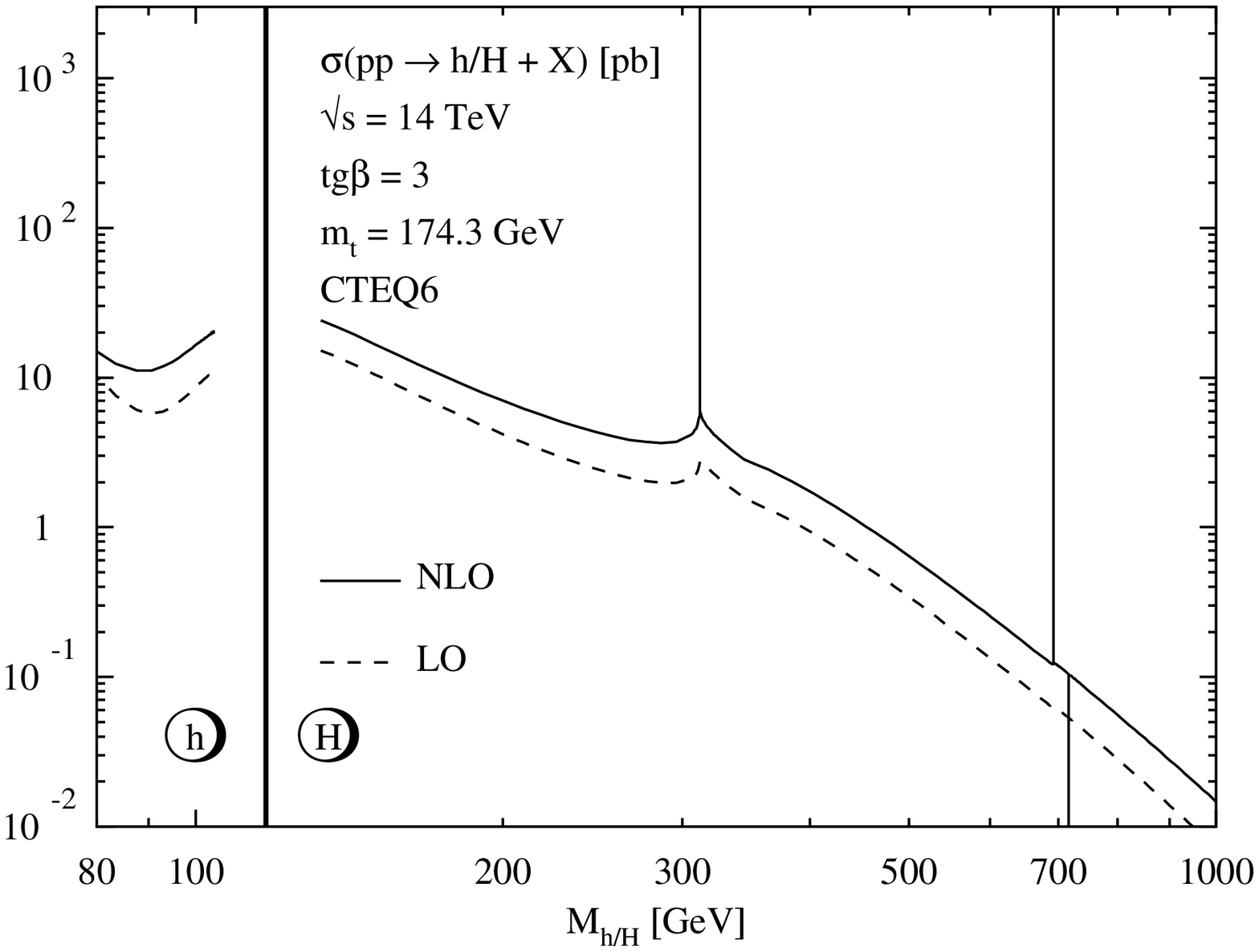}}
\put(40.0,-135.0){\includegraphics{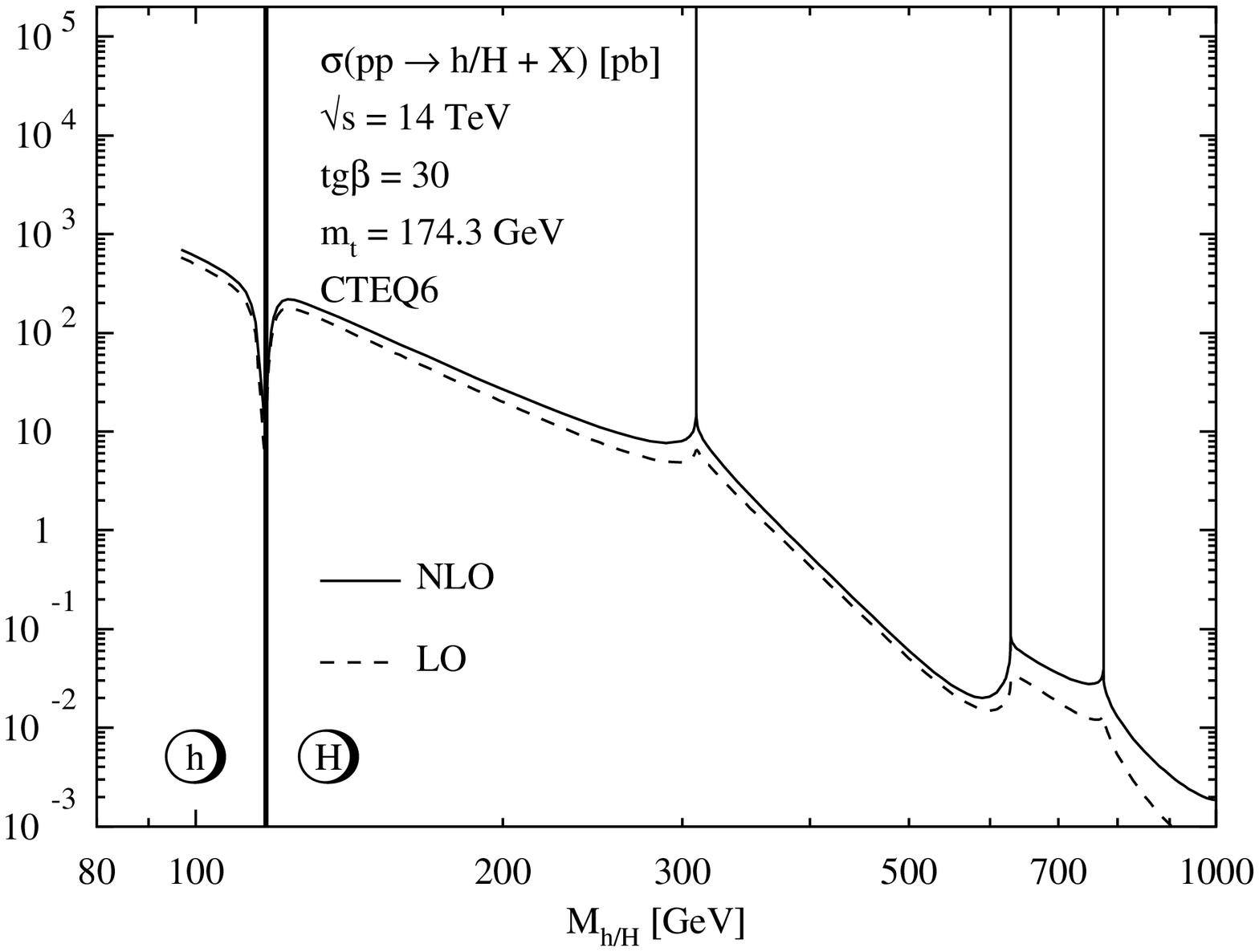}}
\end{picture}
\caption[]{\label{fg:gghnlo} \it Production cross sections of the scalar
MSSM Higgs bosons via gluon fusion as functions of the corresponding
Higgs masses for $\tgb=3$ and 30. The full curves include the QCD
corrections, while the dashed lines correspond to the leading-order
predictions.  The kinks and spikes correspond to the $\tilde
t_1\bar{\tilde t}_1, \tilde b_1\bar{\tilde b}_1$ and $\tilde
b_2\bar{\tilde b}_2$ thresholds in consecutive order with rising Higgs
mass. The renormalization and factorization scales are chosen as the
corresponding Higgs mass.}
\end{figure}

The LO and NLO cross sections are shown in Fig.~\ref{fg:gghnlo}. The QCD
corrections increase the gluon fusion cross sections by 10--100\%, but
can be significantly larger in regions of large destructive
interferences between quark and squark loops, as is the case for very
large Higgs masses for $\tgb=30$. The corrections are of very similar
size for the quark and squark loops individually in agreement with the
results of Ref.~\cite{gghnlosq}. In spite of the large corrections the
residual scale dependence is reduced from about 50\% at LO to $\sim
20\%$ at NLO and indicates a significant stabilization of the theoretical
predictions. This agrees with the former results for the quark loop
contributions \cite{gghnlo,gghnlolim}. Based on the approximate NNLO and
NNNLO results in the limit of heavy top quarks a further moderate
increase by less than 20--30\% can be expected beyond NLO for small and
moderate values of $\tgb$. For large
values of $\tgb$, however, the size of the NLO corrections is moderate
in regions without strong destructive interference effects between the
quark and squark loops
and the scale dependence is small. This signalizes a much more reliable
result after including the NLO corrections. The residual theoretical
uncertainties of our NLO results can be estimated to less than about 20\%.

\begin{figure}[hbtp]
\begin{picture}(100,500)(0,0)
\put(40.0,120.0){\includegraphics{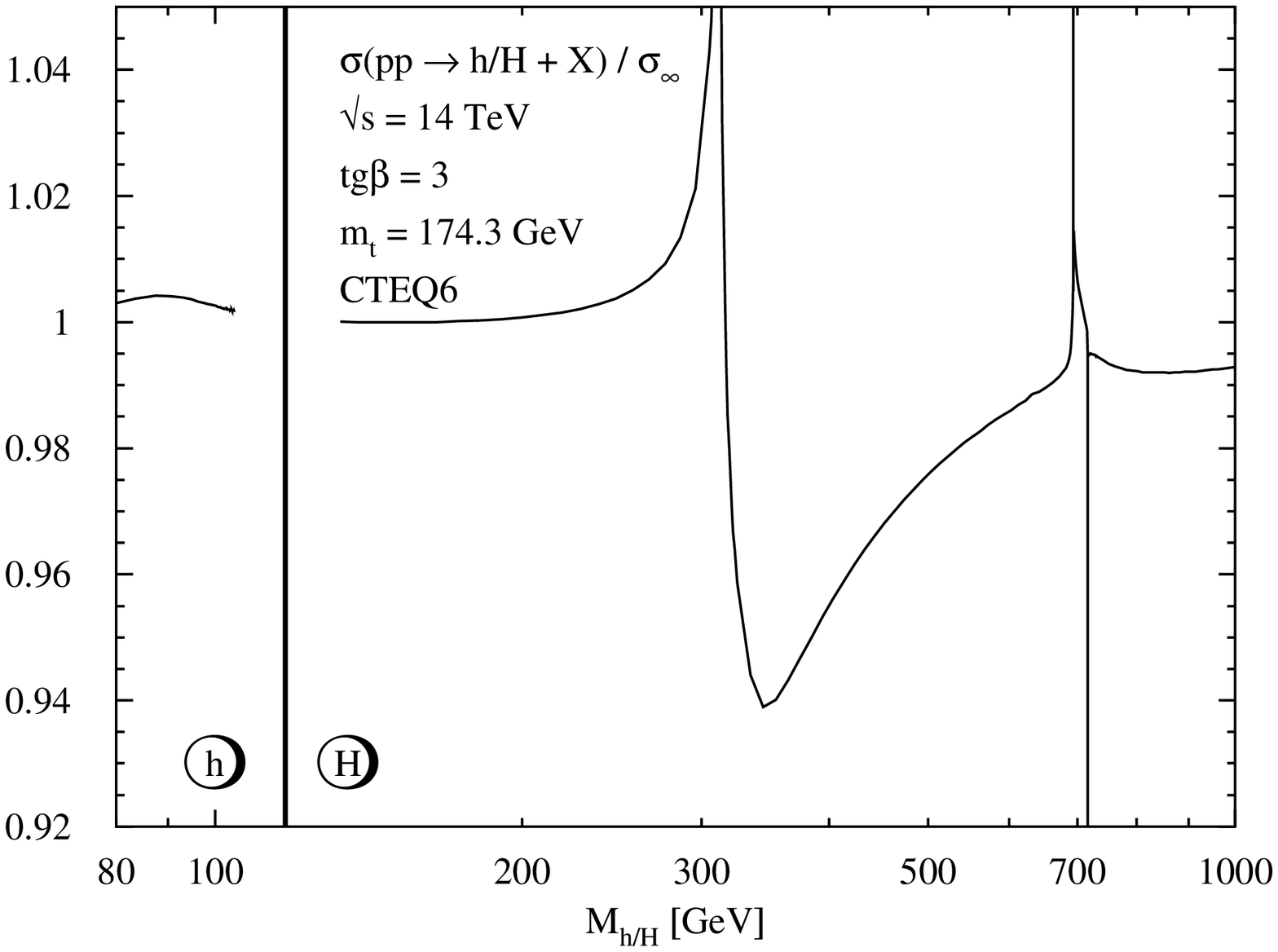}}
\put(40.0,-135.0){\includegraphics{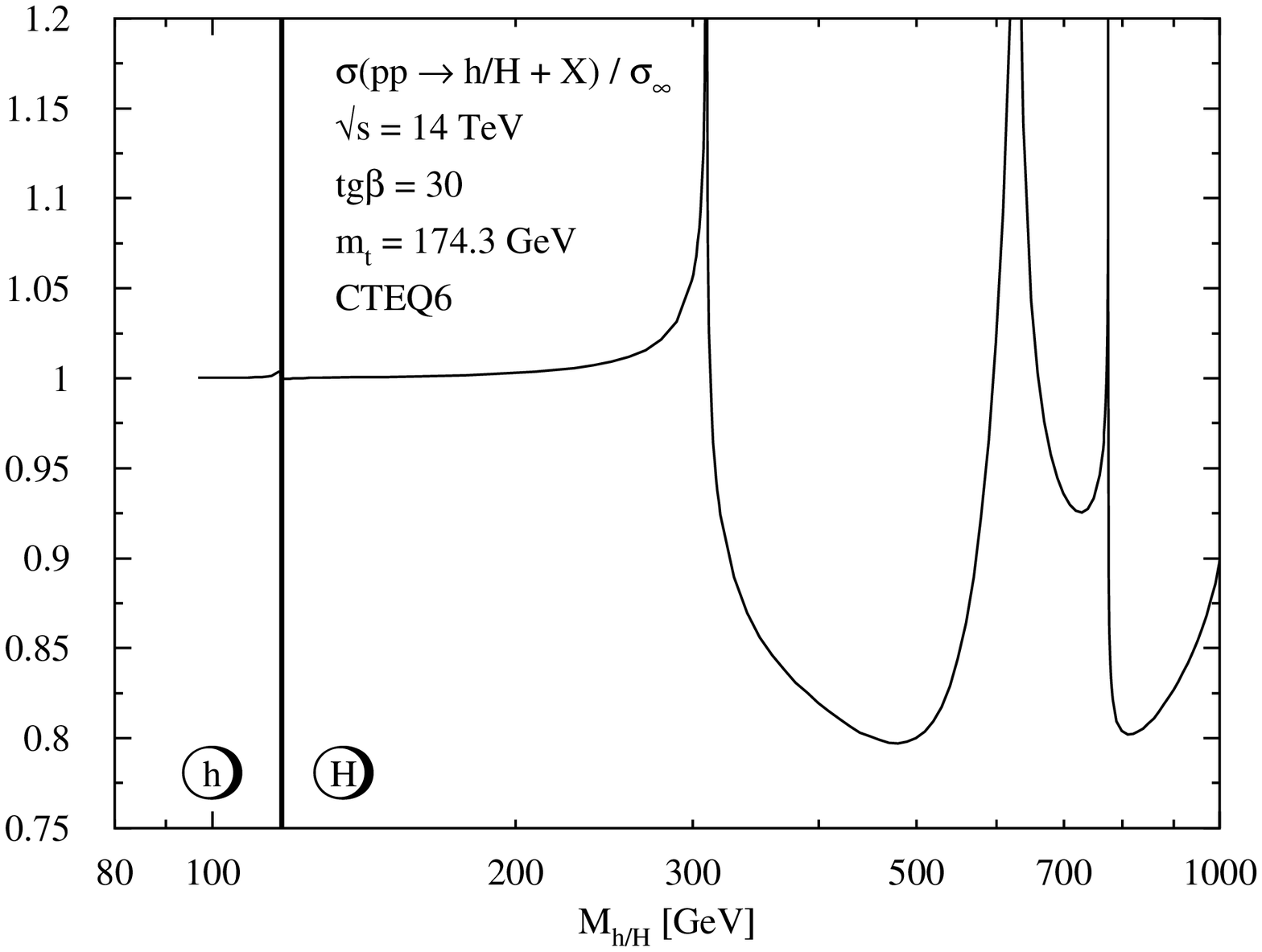}}
\end{picture}
\caption[]{\label{fg:gghmass} \it Ratio of the QCD corrected production
cross sections of the scalar MSSM Higgs bosons via gluon fusion
including the full squark mass dependence and those obtained by taking
the relative QCD corrections to the squark loops in the heavy mass limit
as functions of the corresponding Higgs masses for $\tgb=3$ and 30.  The
kinks and spikes correspond to the $\tilde t_1\bar{\tilde t}_1, \tilde
b_1\bar{\tilde b}_1$ and $\tilde b_2\bar{\tilde b}_2$ thresholds in
consecutive order with rising Higgs mass.  The renormalization and
factorization scales are chosen as the corresponding Higgs mass.}
\end{figure}

The squark mass effects on the $K$ factors are exemplified in
Fig.~\ref{fg:gghmass}, where the ratios of the NLO cross sections are
displayed, including the full mass dependence, and of the NLO cross
sections, where the coefficients $c^{h/H}(\tau_Q,\tau_{\tilde Q})$ and
$d^{h/H}_{ij}(z,\tau_Q,\tau_{\tilde Q})$ are used in the heavy squark
limits ($\tau_{\tilde Q}\to \infty$). In addition to the squark mass
dependence of the LO cross sections, the $K$ factors develop a squark
mass dependence of up to about 20\%, thus supporting the relevance of our
results compared to the previous results of Ref.~\cite{gghnlosq}. The
squark mass effects on the $K$ factors turn out to be larger than the
corresponding quark mass effects \cite{limit}. In addition they are
larger than the residual theoretical uncertainties and cannot be
neglected in realistic analyses. Since the gluino contributions are
expected to be much smaller, the squark mass dependence
obtained in this work will be the dominant part of the differences between
the heavy mass limits and a full MSSM calculation at NLO.


\section{Conclusions}
We have presented a NLO QCD calculation for the squark loop
contributions to the production of neutral scalar MSSM Higgs bosons at
the LHC and their decay modes into gluons and photons. The photonic
couplings of these Higgs particles play a significant role at a future
photon collider, which may be built by Compton backscattering of laser
light from electron beams of a linear $e^+e^-$ collider. The corrections
stabilize the theoretical predictions compared to the LO predictions.
The QCD corrections turn out to be sizeable for the photonic Higgs
couplings and large for the gluonic Higgs decays as well as the gluon
fusion processes.  They increase the latter cross sections and gluonic
decay widths significantly so that the results have to be taken into
account for reliable analyses based on these processes.  Despite of the
large size of the NLO corrections the approximate results in the heavy
top mass limits for the gluonic decay widths and gluon-fusion cross
sections indicate sufficient perturbative convergence. The corrections
beyond NLO are expected to be of moderate size for the central scale
choices. Squark mass effects on the relative QCD corrections are
sizeable and larger than the corresponding quark mass effects for the
quark loop contributions. It should be noted that squark mass effects
are always relevant, if the squark masses are of the order of the top
mass, or the Higgs mass is larger than the corresponding virtual
squark-antisquark threshold. Our results have been implemented in the
program HIGLU \cite{higlu} and can thus be applied to other MSSM
scenarios, too. \\

\noindent
{\bf Note added in proof.} During the final write-up of our work two
independent papers appeared \cite{concur,concur1}, where the virtual
corrections to the quark and squark loops of the gluon fusion processes
$gg\to h/H$ have been derived analytically. However, a full numerical
analysis of the gluon fusion processes at NLO is not contained in these
papers. We have compared our virtual corrections with the authors of
Ref.~\cite{concur1} and found full numerical agreement for the virtual
corrections to the photonic and gluonic Higgs couplings. \\

\noindent
{\bf Acknowledgements.}
We would like to thank P.M.~Zerwas for valuable comments on the
manuscript. M.S. would like to thank LAPTH for their very kind
hospitality during his stay, where major parts of this work have been
performed. We are grateful to R.~Bonciani for the detailed comparison of
the partial results of Ref.~\cite{concur1} with ours.

\end{document}